\documentclass[aps, twocolumn, showpacs, superscriptaddress,groupedaddress, nofootbib, floatfix]{revtex4-1}  
\usepackage{graphicx}  
\usepackage{dcolumn}   
\usepackage{bm}        
\usepackage{amssymb}
\usepackage{amsmath}   
\usepackage{hyperref}
\hypersetup{colorlinks,linkcolor={blue},citecolor={blue},urlcolor={red}}
\usepackage{float}

\begin{document}
\title{Note on stability in conformally connected frames}

\author{Debottam Nandi}
\affiliation{Department of Physics, Indian Institute of
	Technology Madras, Chennai 600036, India}

\email{debottam@physics.iitm.ac.in}

	
	\begin{abstract}
		Since conformal transformations are metric field reparametrization, dynamics in conformally connected frames are often referred to as `equivalent'. In the context of cosmology, while the perturbations remain invariant for a single scalar field model, the background equations differ and therefore the dynamics. However, since the background dynamics are not the same, it is not clear whether the attractor nature of the solutions remains the same in all conformally connected frames, i.e., a stable solution in one frame implies an equivalent stable solution in another frame. To answer the question, in this work, we first consider power law cosmology in the Brans-Dicke theory as well as in the conformal Einstein frame. We show that, in this case, the attractor behavior is indeed equivalent under conformal transformation, i.e., an attractor solution in one frame implies an attractor solution in another frame. However, the decay rates of the deviations from the fixed points are different in the two frames. We are able to relate the behavior and find that the difference is due to the difference in e-fold `clocks' in different frames, i.e., $\Delta N$ in different frames differ from one to another. We show that the behavior is indeed true for any model in conformally connected frames and obtain the general `equivalence' relation. In the context of inflation, we consider two models: Starobinsky and chaotic inflation, and explicitly point out the differences in these two frames. We show that the duration of inflation in any model in the Jordan frame is always higher than the Einstein frame due to the same reason.
	\end{abstract}

\maketitle

\section{Introduction}
In solving non-linear differential equations, initial conditions play a crucial role in determining the dynamical behavior of the system as different initial conditions may lead to different trajectories which may behave completely differently. While most of the non-linear solutions depend highly on initial conditions, there exist some solutions which are independent of the initial conditions. These are referred to as {\emph{attarctors}}.

For simplicity, it is assumed that the early Universe was driven by a scalar field(s). 
Since the corresponding field equations are highly non-linear in nature, it possesses the same problem of arbitrariness of the initial conditions as there is no such reason for choosing precise initial conditions.

Currently, inflationary paradigm \cite{STAROBINSKY198099, Sato:1981, Guth:1981, LINDE1982389, Albrecht-Steinhardt:1982, Linde:1983gd, Mukhanov:1981xt, HAWKING1982295, STAROBINSKY1982175, Guth:1982, VILENKIN1983527, Bardeen:1983, Starobinsky:1979ty} is the most successful paradigm that not only satisfies all observational constraints but also solves horizon and flatness problem. However, perhaps the greatest achievement of the inflationary paradigm is that most models in this paradigm solve the initial value problem, i.e., the solutions are attractors and independent of the initial conditions. Unfortunately, even with tighter observational constraints, ruling out a significant number of models in inflationary paradigm is still difficult and this is the main drawback of this paradigm \cite{Martin:2010hh, Martin:2013tda, Martin:2013nzq,  Martin:2014rqa, Gubitosi:2015pba}. Due to the problems with the inflationary paradigm,  there is a growing interest in finding new alternatives to inflation. Amongst them, classical bouncing models are the most popular one \cite{Wands:1998yp, Novello:2008ra,Cai:2014bea, Battefeld:2014uga, Lilley:2015ksa, Ijjas:2015hcc, Brandenberger:2016vhg}. However, the problem with most bouncing models is that, either the models are extremely difficult to construct using a single scalar field (e.g., ekpyrotic model as the single scalar field action does not lead to scale-invariant spectra) or they behave as non-attractors (e.g., matter bounce).

Inflation or any other alternatives to inflation can be studied in both minimal Einstein frame as well as in non-minimal frame \cite{Horndeski1974, Deffayet2009, Nojiri:2010wj, Nojiri:2017ncd}. The non-minimal coupling to gravity entails a coupling between the curvature and the scalar field, which mixes the metric and scalar degrees of freedom. This also implies that the effective Planck mass during the early Universe is field dependent, and hence, time dependent. The frame is called Jordan where the coupling is the simplest, i.e., scalar field couples with Ricci scalar. The simplest Jordan frame theory is Brans-Dicke \cite{Brans-Dicke1961} where the field is also canonical.

The Jordan-type non-minimal coupling can easily be removed and the theory can be mapped to minimal Einstein theory by performing a suitable conformal transformation. The conformal transformation can be
seen as a metric redefinition. It has been shown that, while the background equations in two frames differ and they are related by the conformal transformation for a single field theory, curvature and tensor perturbations remain invariant \cite{Chiba:2008ia, Gong:2011qe, Kubota:2011re, Weenink:2010rr, Prokopec:2013zya, Karam:2017zno, Chakraborty:2018ost, Karam:2019dlv}. However, differences in the dynamics in these two frames has been pointed out in several articles in the literature \cite{Steinwachs:2013tr, Bezrukov:2009db, DeSimone:2008ei, Barvinsky:2008ia, Capozziello:2006dj, Briscese:2006xu, White:2012ya, White:2013ufa, Nojiri:2000ja, Bahamonde:2016wmz, Bahamonde:2017kbs}.

The motivation of our work is to find out the difference between the two frames in the context of early Universe scenario, mainly inflation. The aim is to study  and compare the attractor behavior of the solutions in these two frames. Since background equations are equivalent (not identical, but related by conformal transformation), we expect equivalent nature of the attractor behavior, too, in both frames. However, in our previous work \cite{Nandi:2018ooh}, we found that, while in the minimal Einstein frame, the contracting (bounce) solutions of the Universe are in general non-attractors; in a simple non-minimal frame, this difficulty can be avoided and we can construct models which lead to a stable contracting solutions of the Universe. This leads to the following questions:
\begin{itemize}
	\item[(a)] Does attractor solution in one frame implies that the equivalent solution in other frame is also an attractor?
	\item[(b)] If they are not the same, why is it?
	\item[(c)] If they are the same, what is the relation between them? Are they identical?
\end{itemize}

It is very difficult to distinguish the physics in these two frames in the context of the inflationary scenario. Therefore, if the attractor behavior also becomes identical in these two frames, then from the current observational framework, it is very difficult to distinguish between the two frames. However, if they differ, then the reheating physics may be different in different frames as (a) it generally arises due to additional relativistic fluid, which in return, may depend on the attracting behavior of the model, and (b) difference in power spectra (mainly $n_s$ and $r$) in two frames may not seen in the CMB but can change the reheating temperature. 

In doing so, we concentrate on the simplest Jordan frame model: Brans-Dicke theory. The detailed phase space analysis in the Brans-Dicke theory has been studied in Refs. \cite{Kolitch:1994kr, Kolitch:1994qa, Santos:1996jc, Holden:1998qg, Hrycyna:2013hla, Hrycyna:2013yia, Kofinas:2016fcp, Nandi:2018ooh}. In a similar manner, also in this work, we will first consider the consequences of power law cosmology for \emph{simplicity} and \emph{better understanding,} and then we will focus on the more realistic inflationary models. In doing so, we will mainly study $f(R)$ theories as the theory is equivalent to Brans-Dicke theory with Brans-Dicke parameter $\omega_{\rm BD} = 0$ \cite{Whitt:1984pd, Muller:1987hp, BARROW1988515, Maeda:1988ab, Ketov:2012jt, 0022-3689-5-6-005, DeFelice:2010aj, Capozziello:2010zz}. We will qualitatively show that the stability of the all conformally connected solutions is indeed equivalent, i.e., a stable solution in one frame ensures the stability of the conformal solution in another frame. However, there is a difference in attractor behavior in these frames and they are closely related to the difference in `e-fold' clocks in these two frames. In the context of slow-roll inflation, we will consider Starobinsky \cite{STAROBINSKY198099, STAROBINSKY1982175} and Chaotic inflation \cite{Linde:1983gd}, and explicitly show the differences of each model in these two frames.

The work is organized in the following way. In the next section, we briefly introduce the Brans-Dicke theory and the corresponding conformal theory in the minimal Einstein frame. In section \ref{Sec:PowerLawCosmo}, we formulate the power law scale factor solutions as well as potentials in both frames. We introduce the phase space in both frames in section \ref{Sec:PhaseSpace} and analyze the phase space and obtain the fixed points as well as the stability condition in section \ref{Sec:FixedPoints}. In section \ref{Sec:Equivalence}, we study the equivalence of the stability of the fixed points and in section \ref{Sec:Attractiveness}, we establish the relation of the attractor behavior in two frames. In the following section \ref{Sec:InflationaryContext}, we focus on the inflationary paradigm and consider two popular inflationary models: Starobinsky and chaotic inflation and compare the attractor behavior between the two frames. In section \ref{Sec:e-folds}, we show the difference in number e-folds in these inflationary models between two frames. In the end, we conclude our work with the future outlook.

In this work, we use the $(-, +, +, +)$ metric signature convention. $\nabla_\mu$ and $\Box \equiv g^{\mu \nu} \nabla_{\mu \nu}$ are defined as the covariant derivative and the d'Alembertian operator in curved spacetime, respectively.  An overdot is defined as the partial derivative with respect to the cosmic time, $\frac{\partial }{\partial t}$. Also, all physical quantities in the Einstein frame are denoted with the tilde, e.g., $\tilde{X}$.

\section{Governing equations in conformally connected frames}\label{Sec:GovEqs}

In this section, we briefly introduce the Brans-Dicke theory \cite{Brans-Dicke1961} as well as the corresponding conformal theory in the Einstein frame. Brans-Dicke theory is the simplest theory in the Jordan frame, i.e., curvature scalar is non-minimally coupled to a canonical scalar field. However, in the Einstein frame, the conformal transformation makes the field non-canonical. In order to retrieve the canonical form of the scalar field, we need another scalar field redefinition. In this section, we provide the action as well as the governing equations in these two frames and the relations between them.

\subsection{Non-minimal frame: Brans-Dicke theory}\label{SubSec:Brans}

The action for the Brans-Dicke theory \cite{Brans-Dicke1961} is

\begin{eqnarray}\label{Eq:ActionBDGeneral}
\mathcal{S}_{\text{BD}} = \frac{1}{2}\int {\rm d}^4 {\bf x} \,\sqrt{-g}\Big(\varphi R - \frac{\omega_{\rm BD}}{\varphi}\,g^{\mu \nu} \partial_\mu \varphi \partial_\nu \varphi  \nonumber\\
- 2\,V(\varphi)\Big),
\end{eqnarray}
where, $R$ is the Ricci scalar, $\omega_{\rm BD}$ is the Brans-Dicke parameter, $\varphi$ is the scalar field and $V(\varphi)$ is the corresponding scalar field potential. Varying the action (\ref{Eq:ActionBDGeneral}) with respect to the metric $g_{\mu \nu}$ provides the metric field equation as

\begin{eqnarray}\label{Eq:EoMBDGengmunu}
\varphi\left(R_{\mu \nu} - \frac{1}{2}g_{\mu \nu}R\right) = \frac{\omega_{\rm BD}}{\varphi} \Big(\nabla_\mu \varphi \nabla_\nu \varphi - \frac{1}{2} g_{\mu \nu} \nabla_\alpha \varphi \nabla^\alpha \varphi \Big) \nonumber\\
- g_{\mu \nu} V(\varphi)  -\left(g_{\mu \nu }\,\Box\varphi  - \nabla_{\mu \nu}\varphi \right).\nonumber\\
\end{eqnarray}
If we take the trace of this equation (\ref{Eq:EoMBDGengmunu}), it becomes
\begin{eqnarray}\label{Eq:EoMBDGenTrace}
R  = \frac{\omega_{\rm BD}}{\varphi^2} g^{\alpha\beta}\nabla_\alpha\varphi\nabla_\beta\varphi + \frac{4}{\varphi}V(\varphi) + \frac{3}{\varphi}\Box\varphi.
\end{eqnarray}
Also, variation of the action (\ref{Eq:ActionBDGeneral}) with respect to the scalar field gives
\begin{eqnarray}\label{Eq:EoMBDGenScalarField}
\Box\varphi  = \frac{1}{2\varphi} g^{\alpha\beta} \nabla_\alpha\varphi\nabla_\beta\varphi - \frac{\varphi}{2 \omega_{\rm BD}}\left(R - 2V_{\varphi}(\varphi)\right),
\end{eqnarray}

\noindent where, $V_\varphi$ is defined as the partial derivative with respect to the scalar field, i.e.,  $\partial V(\varphi)/\partial\varphi$. These three equations (\ref{Eq:EoMBDGengmunu}), (\ref{Eq:EoMBDGenTrace}) and (\ref{Eq:EoMBDGenScalarField}) are not independent; only two of them are independent.

We can express the above equations in terms of homogeneous and isotropic background line element, also known as the the Friedmann-Lema\^\i tre-Robertson-Walker (FLRW) line element:
\begin{eqnarray}\label{Eq:FRWLineElement}
{\rm d}s^2 = - {\rm d}t^2 + a^2(t)\,{\rm d {\bf x}^2},
\end{eqnarray}
where $t$ is the cosmic time and $a(t)$ is the scale factor of the Universe. Using the above metric, the equations (\ref{Eq:EoMBDGengmunu}), (\ref{Eq:EoMBDGenTrace}) and (\ref{Eq:EoMBDGenScalarField}) can be reduced further. The 0-0 component of the equation (\ref{Eq:EoMBDGengmunu}) becomes the energy constraint equation with the form
\begin{eqnarray}\label{Eq:0-0BackBD}
3H^2 = \frac{\omega_{\rm BD}}{2}\frac{\dot{\varphi}^2}{\varphi^2} + \frac{V(\varphi)}{\varphi} - 3H\frac{\dot{\varphi}}{\varphi},
\end{eqnarray}
where, $H(t) \equiv \dot{a}(t)/a(t)$ is the Hubble parameter. The trace equation (\ref{Eq:EoMBDGenTrace}) in FLRW Universe becomes the acceleration equation
\begin{eqnarray}\label{Eq:AccBackBD}
\dot{H} = - \frac{\omega_{\rm BD}}{2}\frac{\dot{\varphi}^2}{\varphi^2} - \frac{1}{3 + 2 \omega_{\rm BD}}\frac{2 V(\varphi) - \varphi V_\varphi(\varphi)}{\varphi} + 2 H \frac{\dot{\varphi}}{\varphi}\nonumber\\
\end{eqnarray}
and the scalar field equation (\ref{Eq:EoMBDGenScalarField}) becomes
\begin{eqnarray}\label{Eq:ScBackBD}
\ddot{\varphi} + 3 H \dot{\varphi}  = 2 \frac{2 V(\varphi) - \varphi V_\varphi(\varphi)}{3 + 2 \omega_{\rm BD}}.
\end{eqnarray}

\subsection{Conformal Einstein frame}\label{Sec:ConformalEinstein}

The appropriate metric transformation to map non-minimally coupled Brans-Dicke theory to a minimally coupled Einstein theory is given by

\begin{eqnarray}\label{Eq:ConfTrans}
	\tilde{g}_{\mu \nu} = \varphi \,g_{\mu \nu}.
\end{eqnarray}

\noindent $\tilde{g}_{\mu \nu}$ is the metric in the Einstein frame. Under the above conformal transformation, the square root of the determinant of the metric and the Ricci scalar behave as
 
 \begin{eqnarray}\label{Eq:ConfQuants}
 	\sqrt{-g} &=& \varphi^{-2}\,\sqrt{-\tilde{g}},\\
 	 R &=& \varphi\left(\tilde{R} + 3 \frac{\tilde{\Box}\varphi}{\varphi} - \frac{9}{2}\frac{\tilde{\nabla}_\mu\varphi \tilde{\nabla}^\mu \varphi}{\varphi^2} \right).
 \end{eqnarray}
Since the covariant derivative and the contraction depend on the metric, we use tilde on them, too. Using these relations, we can transform the action (\ref{Eq:ActionBDGeneral}) into
\begin{eqnarray}\label{Eq:ActionEinstein}
	\mathcal{S}_{\rm E} = \frac{1}{2}\int d^4{\rm \bf x} \sqrt{-\tilde{g}}\left(\tilde{R} - \frac{1}{M_{\rm pl}^2}  \tilde{\nabla}_\mu \tilde{\varphi}\tilde{\nabla}^\mu \tilde{\varphi} -\frac{2}{M_{\rm pl}^2} \tilde{V}(\tilde{\varphi})\right),\nonumber\\
\end{eqnarray}
where,
\begin{eqnarray}\label{Eq:ConfEinsField}
	&&\tilde{\varphi} \equiv \sqrt{\frac{3 + 2 \omega_{\rm BD}}{2}}\,M_{\rm pl}\,{\rm ln}\,\varphi,\\
	\label{Eq:ConfEinsPot}
	\mbox{and}\quad&& \tilde{V}(\tilde{\varphi}) \equiv M_{\rm pl}^2 \frac{V(\varphi(\tilde{\varphi}))}{\varphi(\tilde{\varphi})^2}
\end{eqnarray}

\noindent The action (\ref{Eq:ActionEinstein}) is now minimally coupled. Equation (\ref{Eq:ConfEinsField}) is the scalar field redefinition along with the conformal transformation (\ref{Eq:ConfTrans}) to achieve such form in the Einstein frame. The corresponding metric equation as well as the scalar field equations, respectively, are

\begin{eqnarray}
	&&\tilde{R}_{\mu \nu} - \tilde{g}_{\mu \nu} \tilde{R} =\frac{1}{M_{\rm pl}^2} \Bigg(\tilde{\nabla}_\mu \tilde{\varphi} \tilde{\nabla}_\nu \tilde{\varphi} -  \nonumber\\
	&&\qquad\qquad\qquad g_{\mu \nu}\Big(\frac{1}{2} \tilde{\nabla}_\alpha\tilde{\varphi} \tilde{\nabla}^\alpha\tilde{\varphi}+ \tilde{V}(\tilde{\varphi})\Big)\Bigg), \\
	&&\tilde{\Box}\tilde{\varphi} = \tilde{V}_{\tilde{\varphi}} (\tilde{\varphi}).
	\end{eqnarray}
	
These equations in FLRW background (\ref{Eq:FRWLineElement}) become

\begin{eqnarray}\label{Eq:EoMsBackEinst1}
	\tilde{H}^2 &=& \frac{1}{3M_{\rm pl}^2} \left(\frac{1}{2} \dot{\tilde{\varphi}}^2 + \tilde{V}(\tilde{\varphi})\right), \\
	\label{Eq:EoMsBackEinst2}
	 \dot{\tilde{H}} &=& -\frac{1}{2M_{\rm pl}^2} \dot{\tilde{\varphi}}^2,\\
	\label{Eq:EoMsBackEinst3}
	\ddot{\tilde{\varphi}} &+& 3 \tilde{H}\dot{\tilde{\varphi}} + \tilde{V}_{\tilde{\varphi}} = 0.
\end{eqnarray}

\section{Power law cosmology}\label{Sec:PowerLawCosmo}
In order to study the phase space behavior, the simplest way is to consider power law solution of the scale factor. In this case, the equation of state, as well as the slow-roll parameter, become a constant. In this section, we obtain the potential which leads to power law solutions in both frames. In the Einstein frame, the potential directly fixes the scale factor solution, however, in the Brans-Dicke theory, the scale factor cannot be exactly fixed since there is an extra parameter, $\omega_{\rm BD}$. We will use the parameter such that the corresponding Einstein frame scale factor can also be fixed. In this section, we discuss these things in details.

\subsection{Brans-Dicke theory}\label{Sec:PowerlawBransDicke}

As mentioned earlier, amongst (\ref{Eq:0-0BackBD}),(\ref{Eq:AccBackBD}) and (\ref{Eq:ScBackBD}), only two are independent. For simplicity, in this section, we concentrate on power law solution of the scale factor and  we consider the following form:
\begin{eqnarray}\label{Eq:PowerLawScBD}
a(\eta) = a_0 \left(\frac{\eta}{\eta_0}\right)^n, \quad \varphi = \varphi_0\left(\frac{\eta}{\eta_0}\right)^m, \quad V  = V_0 \left(\frac{\eta}{\eta_0}\right)^p,\nonumber\\
\end{eqnarray}
where, $\eta$ is the conformal time and is related to cosmic time by the relation $\eta \equiv \int {\rm d} t/a(t)$. By solving the equations (\ref{Eq:0-0BackBD}),(\ref{Eq:AccBackBD}) and (\ref{Eq:ScBackBD}), we obtain the solution as
\begin{eqnarray}\label{Eq:wbdV0PowerLaw}
&& \omega_{\rm BD} = \frac{-m^2 + m + 2 m n + 2 n + 2 n^2}{m^2}, \nonumber\\
&& V_0 = \frac{(-m + m^2 - 2 n + 4 m n + 4 n^2)\, \varphi_0}{2\, a_0^2},\nonumber\\
\end{eqnarray}
with 
\begin{eqnarray}
p = m- 2 - 2n.
\end{eqnarray}
This leads to 
\begin{eqnarray}\label{Eq:PotBD}
V = V_0 \,\left(\frac{\varphi}{\varphi_0}\right)^q, \quad q = 1 - 2\left(\frac{1 + n}{m}\right).
\end{eqnarray}

\noindent Therefore, in the Brans-Dicke theory, power law potential $V \sim \varphi^q$ leads to power law solution of the scale factor. Also, as mentioned before, there are two independent parameters $(n, m)$ that govern the dynamics where $n$ determines the nature of the Universe with the scale factor $a \sim \eta^n$. In our case, instead of using the variable $m$, we choose a new variable $\alpha$ which is defined as

\begin{eqnarray}\label{Eq:IntroAlpha}
m = 2\,(\alpha - n),
\end{eqnarray}
where $\alpha$ represents the exponent of the power law scale factor $a  \propto \eta^\alpha$ in the conformally connected Einstein frame. This can easily be verified as the corresponding conformal transformation (\ref{Eq:ConfTrans}) from Jordan frame to Einstein frame is
\begin{eqnarray}\label{Eq:BDEinsteinSc}
\tilde{g}_{ij} = \varphi\,g_{ij} \quad
\Rightarrow \alpha = n + \frac{m}{2},
\end{eqnarray}
which is same as (\ref{Eq:IntroAlpha}). Replacing $m$ by $\alpha$ in (\ref{Eq:wbdV0PowerLaw}) and (\ref{Eq:PotBD}), we obtain
\begin{eqnarray}\label{Eq:wbdq}
\omega_{\rm BD} = \frac{-3\, (n - \alpha)^2 + \,\alpha(\alpha+ 1)}{2\, (n - \alpha)^2}, \quad q = 1 + \frac{n + 1}{n - \alpha},\quad
\end{eqnarray}
and 

\begin{eqnarray}\label{Eq:PotAlphaBD}
V_0 = \frac{\alpha(2\alpha -1)\varphi_0}{a_0^2}.
\end{eqnarray}

\subsection{Einstein frame}\label{Sec:PowerLawEinstein}

As mentioned earlier, unlike Brans-Dicke theory, in the Einstein frame, there is only one free parameter, i.e., the potential which itself can fix the scale factor. This implies conformal transformation acts in such a way that it absorbs one of two parameters described in the Brans-Dicke theory and the Einstein frame is described fully by one parameter only. In order to show that, we need to construct the potential. Using (\ref{Eq:ConfEinsField}) and (\ref{Eq:ConfEinsPot}), it takes the form

\begin{eqnarray}\label{Eq:PotEinsq}
	\tilde{V}(\tilde{\varphi}) = M_{\rm pl}^2 \tilde{V}_0 \exp\left(-\sqrt{\frac{2}{3 + 2 \omega_{\rm BD}}}(2 - q)\frac{\tilde{\varphi}}{M_{\rm pl}}\right).
\end{eqnarray}
 
\noindent Further, using both relations in (\ref{Eq:wbdq}), we obtain the one parameter potential as
 
\begin{eqnarray}\label{Eq:PotEinsalpha}
 \tilde{V}(\tilde{\varphi}) = M_{\rm pl}^2 \tilde{V}_0 \exp\left(-\sqrt{\frac{2(1 + \alpha)}{\alpha}}\frac{\tilde{\varphi}}{M_{\rm pl}}\right).
\end{eqnarray}
Note that, $n$, which is the exponent of the scale factor solution in the Brans-Dicke frame, is not present in the potential and $\alpha$ determines the dynamics only.  Using the above potential, and using equations (\ref{Eq:EoMsBackEinst1}), (\ref{Eq:EoMsBackEinst2}) and (\ref{Eq:EoMsBackEinst3}), it can be shown that the scale factor in the Einstein frame is 
 \begin{eqnarray}\label{Eq:ScaleEins}
 	\tilde{a}(\eta) = \tilde{a}_0 \left(\frac{\eta}{\eta_0}\right)^\alpha.
 \end{eqnarray}

\noindent The result is identical to the equation (\ref{Eq:BDEinsteinSc}).

\section{Phase space analysis}\label{Sec:PhaseSpace}

Since we obtained the potentials in both frames that lead to power law solutions, we can construct and study the phase space in both frames. An efficient way of constructing such space is to define dimensionless quantities and obtain the dynamical equations for them. However, defining dimensionless quantities in two different frames may differ from one another as the fields are redefined.

\subsection{Brans-Dicke theory}\label{Sec:PhaseSpaceBransDicke}

In Brans-Dicke theory, the dimensionless quantities are defined as

\begin{eqnarray}\label{Eq:DefXY}
x  \equiv \frac{\dot{\varphi}}{H \,\varphi}, \quad y \equiv\frac{1}{H}\sqrt{\frac{V(\varphi)}{3\varphi}}
\end{eqnarray}
where $V(\varphi)$ is given by the relation $(\ref{Eq:PotBD})$. Then, using these quantities, the energy constraint equation (\ref{Eq:0-0BackBD}) becomes a constraint equation in the phase space and  can be written as
\begin{eqnarray}\label{Eq:Constrainedxy}
1 + x - \frac{\omega_{\rm BD} }{6} x^2 - y^2 = 0.
\end{eqnarray}
Similarly, the acceleration equation (\ref{Eq:AccBackBD}) can be expressed as
\begin{eqnarray}\label{Eq:ConstrainedAcc}
\frac{\dot{H}}{H^2} = 2x - \frac{\omega_{\rm BD}}{2}x^2 - \frac{3(2 - q)}{3 + 2 \omega_{\rm BD}}y^2 .
\end{eqnarray}

Defining the equation of state of the system in a non-minimally coupled frame is not unique as the energy momentum tensor is coupled to the other field. Instead, we  re-arrange and re-write field equations in a similar fashion in Einstein frame, i.e., $G_{\mu \nu} = T^{(\rm eff)}_{\mu \nu}(g_{\alpha\beta}, \varphi)$, where $G_{\mu \nu}$ is the Einstein tensor and we can define $T^{(\rm eff)}_{\mu \nu}(g_{\alpha\beta}, \varphi)$ as the effective energy momentum tensor.  This ensures that the definition of the effective equation of state becomes identical to the definition in Einstein frame and it depends only on the scale factor in the respective frame:
\begin{eqnarray}\label{Eq:EffectiveEOS}
w_{\rm eff} = -1 - \frac{2}{3}\frac{\dot{H}}{H^2}.
\end{eqnarray}

\noindent As discussed earlier, due to diffeomorphism invariance, the degrees of freedom of the system is one. Hence, the phase space is one dimensional. In order to study the dynamics, instead of using cosmic or conformal time, we use the `e-fold' time convention with $N \equiv {\rm ln}\left|a(t)\right| = \int H(t){\rm d}t$. Using equations (\ref{Eq:0-0BackBD}), (\ref{Eq:AccBackBD}) and (\ref{Eq:ScBackBD}),  we obtain the evolution equations of $x$ and $y$ as	
\begin{eqnarray}\label{Eq:EoMX}
\frac{{\rm d} x}{{\rm d}N} &=& - 3x - x^2 - x\,\frac{\dot{H}}{H^2} + \frac{6(2 - q)}{ 3 + 2 \omega_{\rm BD}}y^2  , \\
\label{Eq:EoMY}
\frac{{\rm d} y}{{\rm d}N} &=& \frac{(q - 1)}{2} x y - y\,\frac{\dot{H}}{H^2} .
\end{eqnarray}

\noindent While the direction of $N$ is positive in an expanding Universe, in case of contracting Universe, $N$ runs in the negative direction. Also, $x$ and $y$ are constrained by the equation (\ref{Eq:Constrainedxy}) and therefore, only one of the above two equations is independent. 

\subsection{Einstein frame}\label{Sec:PhaseSpaceEinstein}
Similar to Brans-Dicke theory, also in Einstein frame, we can define the dimensionless parameters as
\begin{eqnarray}\label{Eq:DefXYtilde}
	\tilde{x}\equiv \frac{\dot{\tilde{\varphi}}}{\sqrt{6} M_{\rm pl} \tilde{H}}, \quad \tilde{y} \equiv \frac{\sqrt{\tilde{V}(\tilde{\varphi})}}{\sqrt{3} M_{\rm pl} \tilde{H}},
\end{eqnarray}
which is different than the same described in (\ref{Eq:DefXY}). $\tilde{V}(\tilde{\varphi})$ is given by the equation (\ref{Eq:PotEinsalpha}). In a similar way, the energy constraint equation in this frame becomes
\begin{eqnarray}
	\tilde{x}^2 + \tilde{y}^2  = 1,
\end{eqnarray}
and the evolution equations of $\tilde{x}$ and $\tilde{y}$ take the form

\begin{eqnarray}\label{Eq:EvotildX}
	&&\tilde{x}^\prime = 3\tilde{x}^3 - 3\tilde{x} + \sqrt{\frac{3(1 + \alpha)}{\alpha}} \,\tilde{y}^2, \\
	\label{Eq:EvotildY}
	&&\tilde{y}^\prime = 3 \tilde{x}^2 \,\tilde{y} - \sqrt{\frac{3(1 + \alpha)}{\alpha}} \tilde{x}\, \tilde{y}.
\end{eqnarray}

\section{Fixed points and stability}\label{Sec:FixedPoints}

Since we constructed the phase space in both frames, now we can obtain the fixed point solutions and study the stability of those points in both frames. The fixed points represent the exact solutions of the system. In order to obtain such points, we need to set the velocities of the phase space variables to be zero (i.e., ${\rm d}x/{\rm d} N = {\rm d}y/{\rm d} N = 0$ and ${\rm d}\tilde{x}/{\rm d} N = {\rm d}\tilde{y}/{\rm d} N = 0$).

\subsection{Brans-Dicke theory}\label{Sec:FixedPointsBransDicke}

Using equations (\ref{Eq:EoMX}) and (\ref{Eq:Constrainedxy}), along with (\ref{Eq:ConstrainedAcc}), we can find the critical or the fixed points of the system in the Brans-Dicke theory. This corresponds to four such points. We are only interested in the first two as the two solutions describe the system as desired, i.e., the scale factor in the Brans-Dicke theory is $a(\eta) \sim \eta^n.$ These are
\begin{eqnarray}\label{Eq:FixedPoint1}
1.\quad x_1^* &=& -2 + \frac{2\alpha}{n}, \quad y_1^* = -  \frac{\sqrt{\alpha(2\alpha-1)}}{\sqrt{3}n} \\
\label{Eq:FixedPoint2}
2. \quad x_2^* &=& -2 + \frac{2\alpha}{n}, \quad y_2^* =   \frac{\sqrt{\alpha(2\alpha-1)}}{\sqrt{3}n}
\end{eqnarray}

\noindent Since $H$ is positive (negative) for the expansion (contraction) of the Universe, the same sign convention in front of $y^*$ defines the expansion (contraction) of the Universe. Therefore, the fixed point (\ref{Eq:FixedPoint1}) is the solution for contracting Universe, whereas, the fixed point (\ref{Eq:FixedPoint2}) represents the expansion of the Universe. The other two fixed points are

\begin{eqnarray}
\label{Eq:FixedPoint3}
3. \quad x_3^* &=& \frac{3 - \sqrt{9 + 6 \omega_{\rm BD}}}{\omega_{\rm BD}}, \quad y_3^* = 0 \\
\label{Eq:FixedPoint4}
4. \quad x_4^* &=& \frac{3 + \sqrt{9 + 6 \omega_{\rm BD}}}{\omega_{\rm BD}}, \quad y_4^* = 0
\end{eqnarray}

These points are the solution for zero potential, i.e., $V_0 = 0$ (whereas, the first two points are valid for non-zero finite potential) and completely governed by the kinetic part of the action (\ref{Eq:ActionBDGeneral}). In this work, we are not interested in these two solutions.

Now, in order to study the stability of these fixed points, we need to linearize any one of the equations (\ref{Eq:EoMX}) and (\ref{Eq:EoMY}) and we can obtain the dynamical solution of the deviations from the fixed points:
\begin{eqnarray}\label{Eq:LinearizedEq}
\delta x(N) = e^{\lambda (N - N_0)}\,\delta x_0, \quad \delta y(N) =\frac{1 - \frac{\omega_{\rm BD}}{3} x^*}{2y^*}\,\delta x(N),\nonumber\\
\end{eqnarray}

\noindent where, $\lambda \equiv \frac{\partial A(x, y(x))}{\partial x}|_*$,  $A(x, y)$ is the right hand side of equation (\ref{Eq:EoMX}) and $|_*$ denotes the value at the fixed point. $\delta x_0$ is the initial value of the deviation from the fixed point. By linearizing equations, we assume that we are studying the stability condition in the vicinity of the fixed points, i.e., the deviations from the fixed points are very small.

A fixed point is stable if $\delta x\,(\mbox{and\,}\delta y)$ approaches zero asymptotically in time, i.e., the deviation vanishes over time. If $\lambda$ is negative (positive) in an expanding (contracting) Universe, then $\delta x$ (and $\delta y$) approach zero as $N$ approaches $\infty\, (-\infty)$. This implies that even if we start from a phase space point which is arbitrarily deviated from the fixed point, the deviated solution eventually returns to the fixed point solution asymptotically in time. This type of solutions are called attractors.

For the fixed points (\ref{Eq:FixedPoint1}) and (\ref{Eq:FixedPoint2}), $\lambda$ takes the form:
\begin{eqnarray}\label{Eq:StabBrans}
	\lambda_{(1,2)} = \frac{1 - 2\alpha}{n}.
\end{eqnarray}

Then, $\lambda$ is positive for $(n > 0,\, \alpha < 1/2)$ and for $(n < 0,\, \alpha > 1/2)$. In this case, the contracting Universe solution (\ref{Eq:FixedPoint1}) is an attractor. Similarly, $\lambda$ is negative for $(n > 0,\, \alpha > 1/2)$ and for $(n < 0,\, \alpha < 1/2)$ and the corresponding expanding Universe solution (\ref{Eq:FixedPoint2}) is an attractor.
\subsection{Einstein frame}\label{Sec:FixedPointsEinstein}

In a similar manner, we can evaluate the fixed points in the Einstein frame. Also in this frame, there are four fixed points. The fixed points corresponding to solution $\tilde{a}(\eta) \sim \eta^\alpha$ are

\begin{eqnarray}\label{Eq:Fixedtilde1}
	\tilde{x}^*_1 &=&  \sqrt{\frac{(1 + \alpha)}{3 \alpha}}, \quad \tilde{y}^*_1 = -\sqrt{\frac{(2 \alpha - 1)}{3 \alpha}}, \\
	\label{Eq:Fixedtilde2}
	\tilde{x}^*_2 &=&  \sqrt{\frac{(1 + \alpha)}{3 \alpha}}, \quad \tilde{y}^*_2 = \sqrt{\frac{(2 \alpha - 1)}{3 \alpha}}.
\end{eqnarray}
Again, the fixed point (\ref{Eq:Fixedtilde1}) is the solution for contraction of the Universe, where as, the fixed point (\ref{Eq:Fixedtilde2}) represents expansion of the Universe. $\tilde{\lambda}$ corresponding to these fixed points are

\begin{eqnarray}\label{Eq:StabEins}
	\tilde{\lambda}_{(1, 2)} = \frac{1- 2 \alpha}{\alpha}.
\end{eqnarray}

\noindent From the above expression, it is obvious that $\tilde{\lambda}$ is positive for $0 <\alpha < 1/2$ and thus the contracting solution (\ref{Eq:Fixedtilde1}) is an attractor solution. For $-\infty < \alpha < 0$ and $\alpha > 1/2$, the expanding Universe solution (\ref{Eq:Fixedtilde2}) is an attractor.

\section{Equivalence of the stability condition}\label{Sec:Equivalence}

By looking at the relations (\ref{Eq:StabBrans}) and (\ref{Eq:StabEins}), note that we can obtain different conditions for stability in conformally connected frames. For example, if we choose $\alpha < -1$ and $n > 0$, then the contracting solution (\ref{Eq:FixedPoint1}) in the Brans-Dicke theory becomes an attractor solution. However, $\alpha < -1$ corresponds to an attractor solution for expanding Universe (\ref{Eq:Fixedtilde2}) in the Einstein frame. Therefore, it seems there is a clear contradiction in stability conditions in two different conformally connected frames.

In order to answer the question, we need the transformation relations between the two dimensionless parameters defined in these two frames, i.e.,  $(x, y)$ and $(\tilde{x}, \tilde{y})$. It can easily be shown that the relations are
\begin{eqnarray}
	x = \frac{\sqrt{6} A \,\tilde{x}}{1 - \sqrt{\frac{3}{2}} A \,\tilde{x}}, \quad y = \frac{\tilde{y}}{1 - \sqrt{\frac{3}{2}} A \,\tilde{x}},
\end{eqnarray}
where,

\begin{eqnarray}
	A \equiv  \sqrt{\frac{2}{3 + 2 \omega_{\rm BD}}}, \quad\mbox{for power law,}\quad \frac{\sqrt{2}(\alpha - n)}{\sqrt{\alpha(\alpha + 1)}}.
\end{eqnarray}

Using the transformation, one can obtain (\ref{Eq:EvotildX}) and (\ref{Eq:EvotildY}) from (\ref{Eq:EoMX}) and (\ref{Eq:EoMY}). In fact, there is no surprise that we can obtain all background equation from one frame to another as this is the conformal transformation in the phase space. Using these relations, one can verify that the fixed points in the Brans-Dicke theory correspond to the desired fixed points in Einstein frame and the relations amongst them are

\begin{eqnarray}\label{Eq:xtildexytildey}
	\tilde{x}^*_{(1, 2)} = \frac{n}{\alpha - n}\,\sqrt{\frac{1 + \alpha}{12 \alpha}}\, x^*_{(1, 2)}, \quad \tilde{y}^*_{(1, 2)} = \frac{n}{\alpha}\,y^*_{(1, 2)}.
\end{eqnarray}

To make things transparent, we take an example: $(n, \alpha) = (2, -3)$. This leads to $\lambda = 7/2$ and $\tilde{\lambda}  = - 7/3$. Therefore, in the Brans-Dicke theory, while the contraction solution (\ref{Eq:FixedPoint1}) is stable, in Einstein frame, the expanding solution (\ref{Eq:Fixedtilde2}) is stable. Since the two frames are connected by a well-defined transformation, there are two possibilities:
\begin{itemize}
	\item[1.] Both frames are expanding and in one frame the solution is stable and in another frame, the solution is not stable. Therefore, the stability is not invariant in two frames, unlike other transformations.
	\item[2.] Stability is invariant and in the Brans-Dicke theory, the Universe is contracting whereas, in the Einstein frame, the Universe is expanding. 
\end{itemize}

Now consider the above relations (\ref{Eq:xtildexytildey}) again. By choosing $(n, \alpha) = (2, -3)$, the relation between $y^*$ and $\tilde{y}^*$ becomes $\tilde{y}^*_{(1,2)} = -2/3\, y^*_{(1,2)}.$  Therefore, in the Brans-Dicke theory, if the Universe is indeed contracting, i.e., $y^*$ is negative and the solution is an attractor, in the Einstein frame, the sign of $\tilde{y}^*$ becomes positive and thus the Universe in the Einstein frame is expanding. Therefore, the solution in the Einstein frame, again, is also an attractor. This can also be verified by looking at the relation between e-folds $N$ and $\tilde{N}$ in two different frames as

\begin{eqnarray}\label{Eq:NtildeN}
	\Delta N  = \frac{n}{\alpha} \,\Delta\tilde{N}
\end{eqnarray}

\noindent   The relation is obtained by using the relation between scale factors in two different frames. Sign of $\Delta N$ determines the direction of $N$, i.e., if $\Delta N$ is positive (negative), the Universe is expanding (contracting).  For $(n, \alpha) = (2, -3),$ $\Delta N = -2/3\,\Delta \tilde{N}$. Hence, in one frame, if the Universe is contracting, i.e., $\Delta N < 0$, in another frame, $\Delta \tilde{N}$ becomes positive and the Universe in that frame is expanding, and vice versa. We can also relate $\lambda$ and $\tilde{\lambda}$ and in case of power law, it becomes

\begin{eqnarray}\label{Eq:ltildel}
\lambda_{(1, 2)} = \frac{\alpha}{n}\,\tilde{\lambda}_{(1,2)}.
\end{eqnarray}

\noindent Using this relation along with equations (\ref{Eq:xtildexytildey}) and (\ref{Eq:NtildeN}), we can also establish the equivalence of the equations (\ref{Eq:LinearizedEq}) in the Einstein frame and it becomes

\begin{eqnarray}\label{Eq:DeltaxEins}
\delta \tilde{x}(\tilde{N}) = e^{\tilde{\lambda} (\tilde{N} - \tilde{N}_0)}\,\delta \tilde{x}_0, \quad \delta \tilde{y}(\tilde{N}) =-\frac{\tilde{x}^*}{\tilde{y}^*}\,\delta \tilde{x}(\tilde{N}).
\end{eqnarray}

It can be verified that the above equations are the solutions for the deviations in the Einstein frame. Therefore, \emph{under conformal transformation, stability is indeed invariant}, i.e., in one frame if the solution is an attractor, in another frame the connected solution is also an attractor and they are appropriately related. This relation can be extended to other fixed points and the corresponding stability of the fixed points as well. It becomes obvious from the example that the stability invariance is indeed true for any model. \emph{This is one of the main results of this work.}


\section{Lyapunov exponent}\label{Sec:Attractiveness}

Since the conformal transformation is not a coordinate transformation, the equivalence of the two frames does not imply the frames are identical as the background solutions differ. Also, as one can see from equation (\ref{Eq:ltildel}), $\lambda$ changes from frame to frame. The parameter $\lambda$, often called the \emph{`Lyapunov exponent',} characterizes the rate of separation between the infinitesimally close fixed point and deviated trajectories, i.e., \emph{how fast the deviation from the stable (unstable) fixed point decays (grows).} However, there is an identity which is invariant under conformal transformation. As we can see from (\ref{Eq:NtildeN}) and (\ref{Eq:ltildel}), 

\begin{eqnarray}\label{Eq:EquivRelAttrac}
	\lambda \, \Delta N = \tilde{\lambda}\,\Delta\tilde{N}
\end{eqnarray}

\noindent  If we consider a model that is not a power law theory, then the individual relation (\ref{Eq:NtildeN}) and (\ref{Eq:ltildel}) may not hold true because of the fact that $\lambda$ can be a function  of $N$ as well. However, \emph{the above identity (\ref{Eq:EquivRelAttrac}) holds true for any model under any conformal transformation.} Therefore, the above relation  indeed represents the stability invariance under any conformal transformation. The equivalent solution of (\ref{Eq:LinearizedEq}) or (\ref{Eq:DeltaxEins}) for any model becomes

\begin{eqnarray}
	\delta x(N) = \exp\left[\int^N_{N_0}\lambda(N)\, {\rm d}N\right]\,\delta x_0.
\end{eqnarray}

\noindent We can explicitly calculate the equivalent relation of (\ref{Eq:NtildeN}) for the general case and it takes the form

\begin{eqnarray}\label{Eq:DeltaN}
\Delta N = \left(1  - \sqrt{\frac{3}{3 + 2 \omega_{\rm BD}}} \tilde{x}\right)\,\Delta \tilde{N}
\end{eqnarray}

\noindent $\tilde{x}$ is the dimensionless variable defined in  (\ref{Eq:DefXYtilde}) in the Einstein frame. This relation tells us that the `e-fold' clocks in different conformally connected frames run differently. Using the above relation and equation (\ref{Eq:EquivRelAttrac}), we obtain the relation between $\lambda$ and $\tilde{\lambda}$ in two conformally connected frames as
\begin{eqnarray}\label{Eq:LambdaFrames}
\lambda = \tilde{\lambda}\left(1  - \sqrt{\frac{3}{3 + 2 \omega_{\rm BD}}} \tilde{x}\right)^{-1}.\quad
\end{eqnarray}

\begin{figure}
\includegraphics[width=\linewidth]{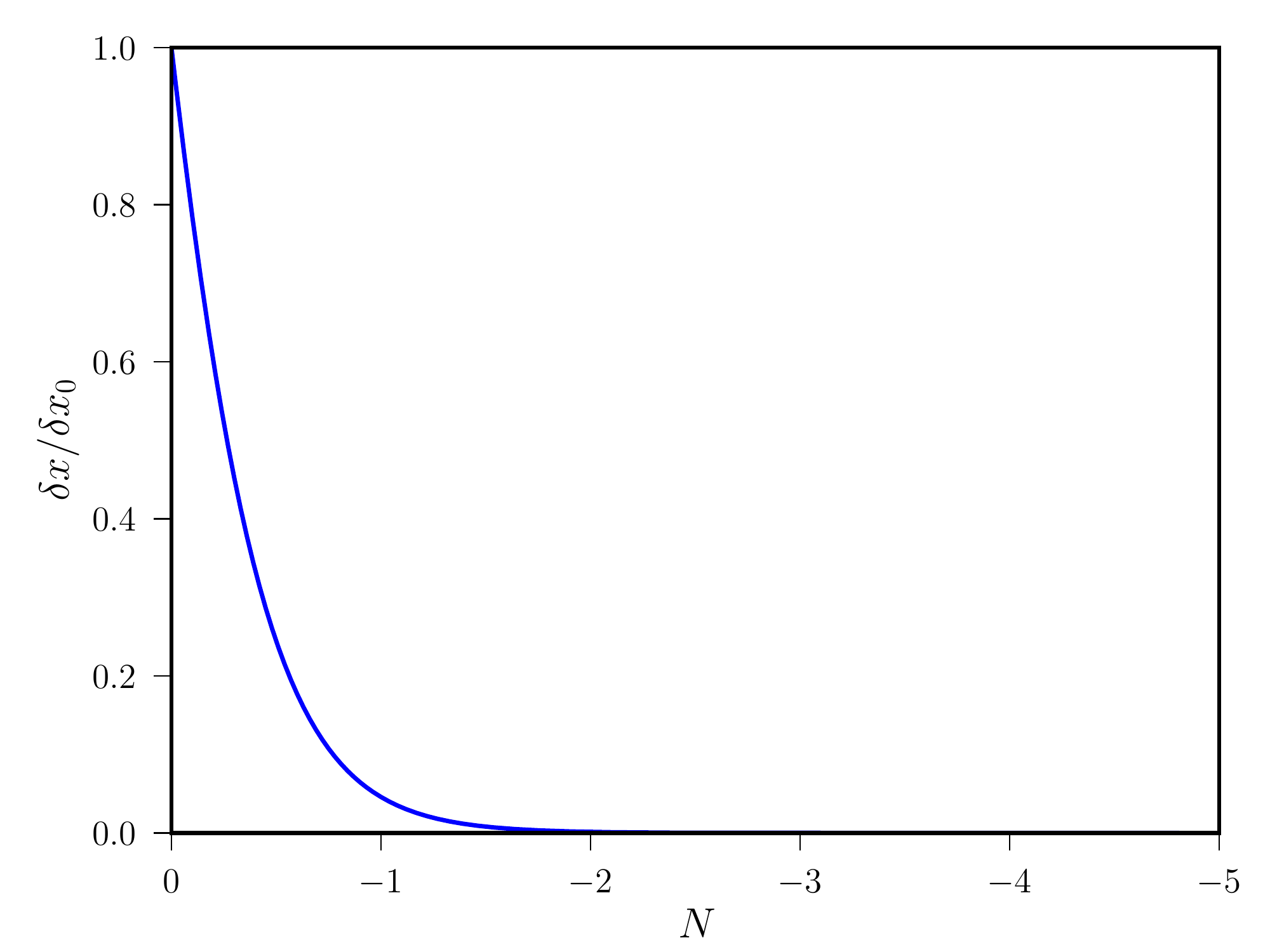}\\
\includegraphics[width=\linewidth]{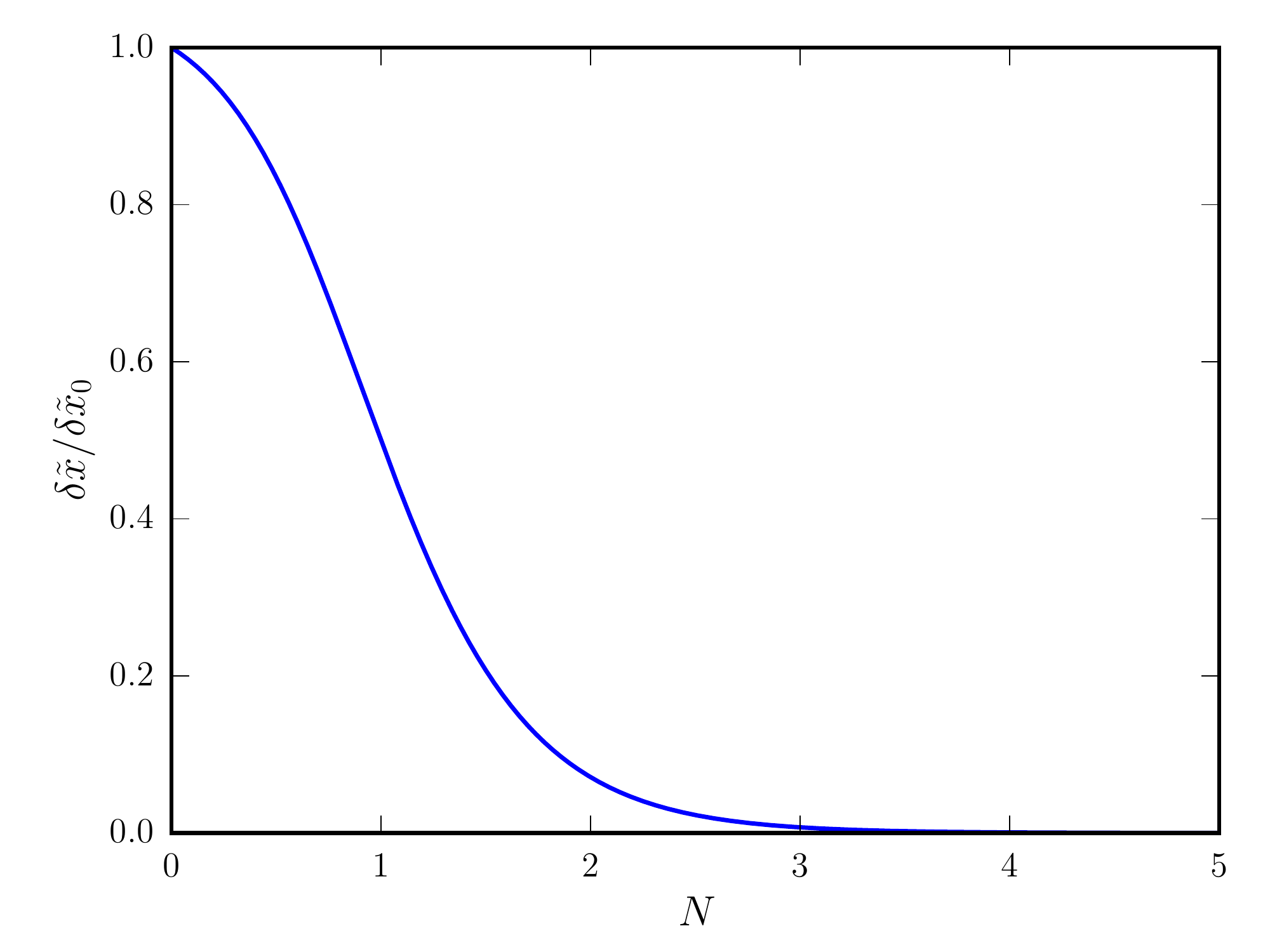}
\caption{At the top, we plot the deviation from the fixed point $x_1 = -5$ in the Brans-Dicke theory. In this frame, The Universe is contracting and therefore, $N$ is decreasing. At the bottom, we plot the deviation from the corresponding fixed point $\tilde{x}_2 = \sqrt{2}/3$ in the Einstein frame. In this frame, the Universe is expanding and hence, $N$ is increasing. The deviations are properly normalized.}
	\label{Fig:devxPowerLaw}
\end{figure}

\noindent Therefore, $\tilde{x} > 0$ implies $\Delta N < \Delta \tilde{N}$ and therefore,   $\lambda > \tilde{\lambda}$. Consider the same example, $(n, \alpha) = (2, -3)$, i.e., in Brans-Dicke theory, the Universe is contracting while in the Einstein frame, the Universe is expanding. From equations (\ref{Eq:FixedPoint1}) and (\ref{Eq:Fixedtilde1}),  the fixed points in the Brans-Dicke theory and the Einstein frame are $x_1^* = -5,\,\tilde{x}_2^* = \sqrt{2}/3$, respectively and $\lambda = -3/2\,\tilde{\lambda}, ~\tilde{\lambda} = 7/3$. Therefore, in the Brans-Dicke theory the deviation decays faster than the same in the Einstein frame. This can be seen in Figure \ref{Fig:devxPowerLaw}.

\begin{figure}
\includegraphics[width=\linewidth]{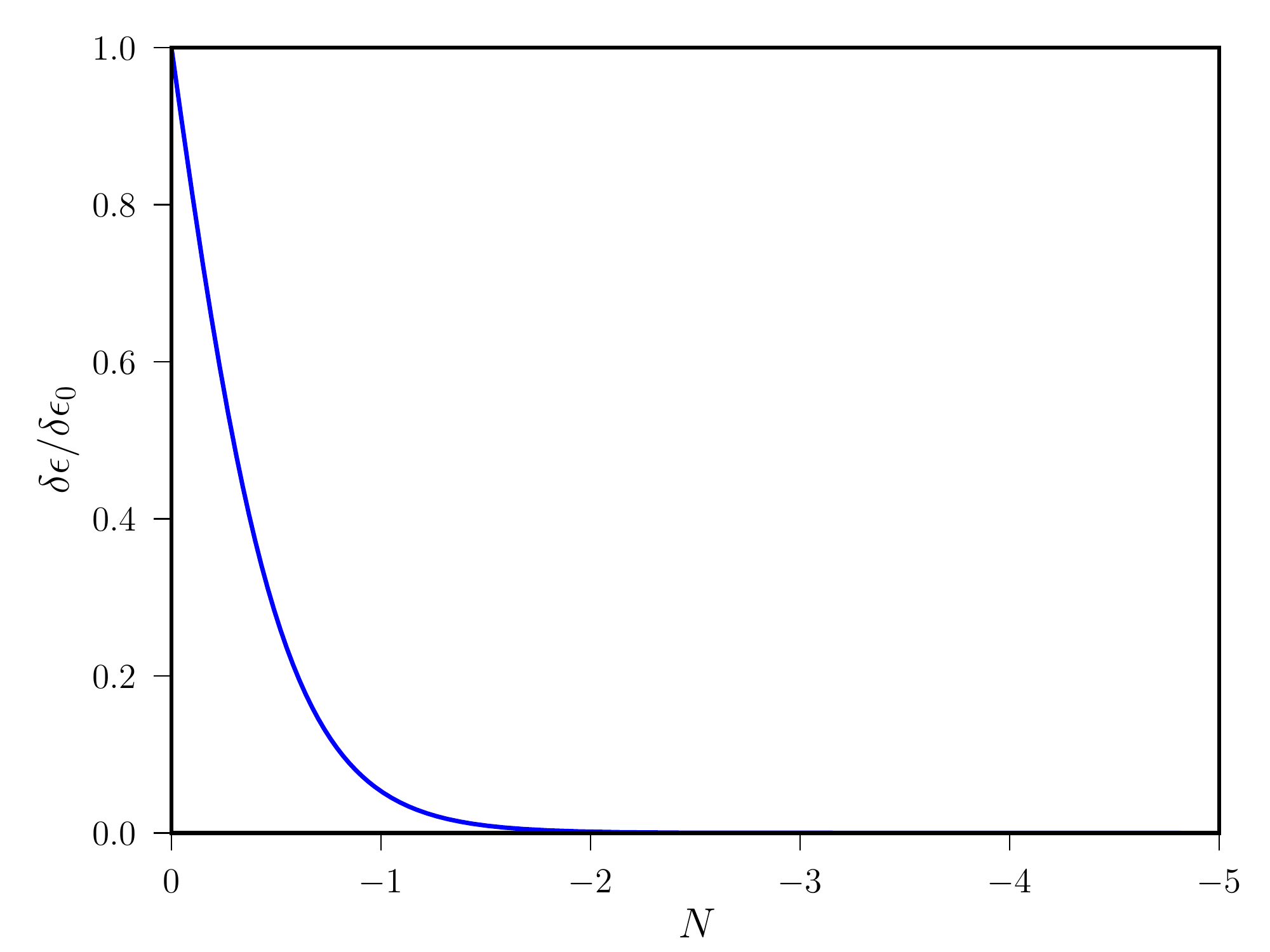}\\
\includegraphics[width=\linewidth]{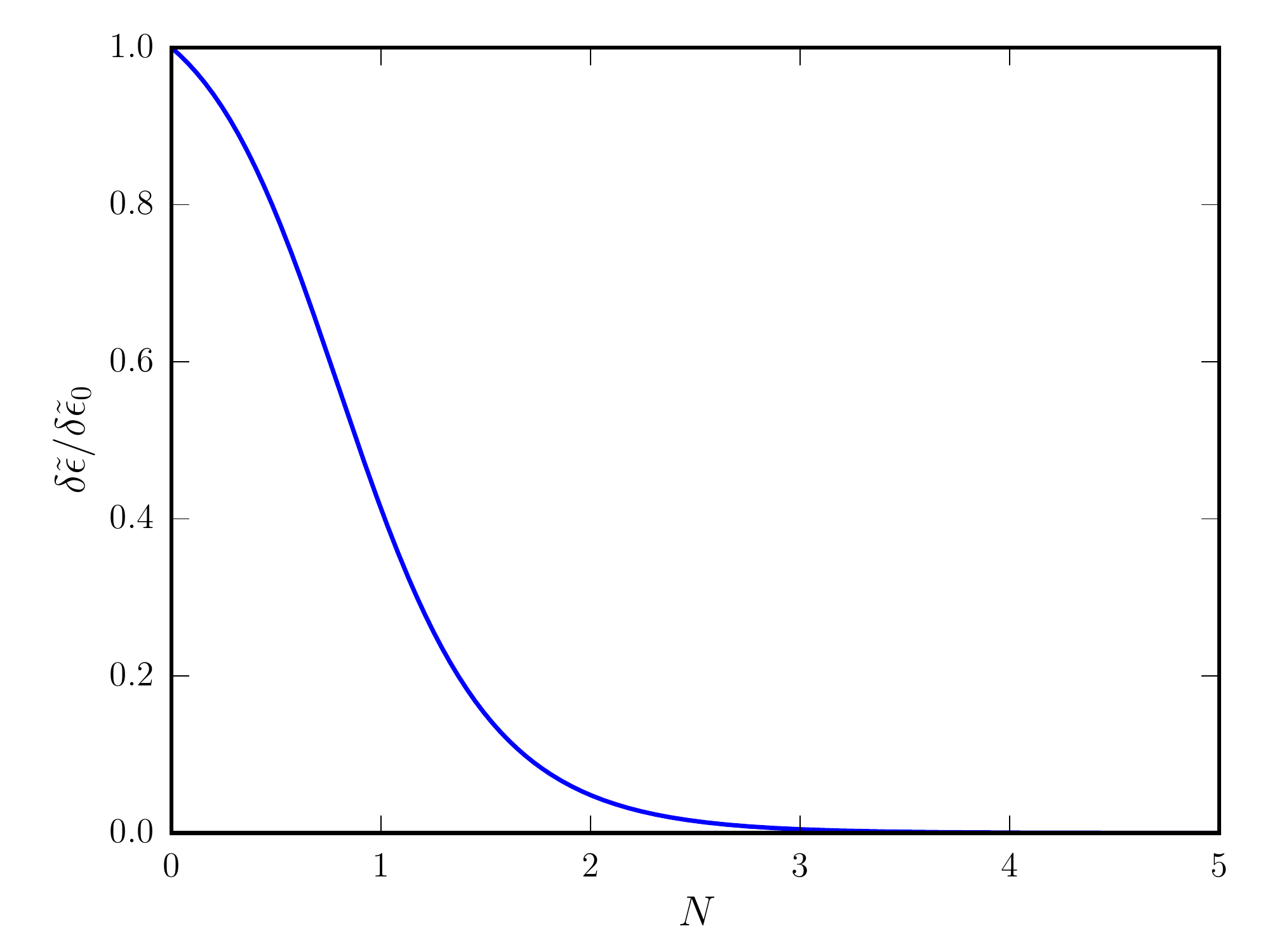}
\caption{At the top, we plot the deviation of the slow-roll parameter for the fixed point $x_1 = -5$ in the Brans-Dicke theory. In this frame, The Universe is contracting and therefore, $N$ is decreasing. At the bottom, we plot the deviation of the slow-roll parameter for the corresponding fixed point $\tilde{x}_2 = \sqrt{2}/3$ in the Einstein frame. In this frame, the Universe is expanding and hence, $N$ is increasing. The deviations are properly normalized.}
	\label{Fig:devslpPowerLaw}
\end{figure}

In Figure \ref{Fig:devxPowerLaw}, at the top, we plot the deviation of  $x$ defined in equation (\ref{Eq:DefXY}) in the Brans-Dicke theory (direction of $N$ is negative) and at the bottom, we plot the deviation of $\tilde{x}$ defined in equation (\ref{Eq:DefXYtilde}) in the Einstein frame (direction of $N$ is positive). The deviations are normalized by dividing $\delta x(N)$ and $\delta \tilde{x}(N)$ by $\delta x_0 \equiv \delta x(0) =  0.5$ and $\delta \tilde{x}_0 \equiv \delta \tilde{x}(0) =  0.5$, i.e., the initial deviations, respectively. As one can easily see, in the Brans-Dicke theory (top), the deviation decays faster than the same in the Einstein frame (bottom). The same is true for the deviation of the slow-roll parameter as well which is plotted in the Figure \ref{Fig:devslpPowerLaw}.

Therefore, under conformal transformations, the attracting behaviors in two different frames differ significantly. In the case of power law solutions, the difference in the attractor behavior depends on the exponents of the scale factors in two different frames.

\begin{figure}
\includegraphics[width= \linewidth]{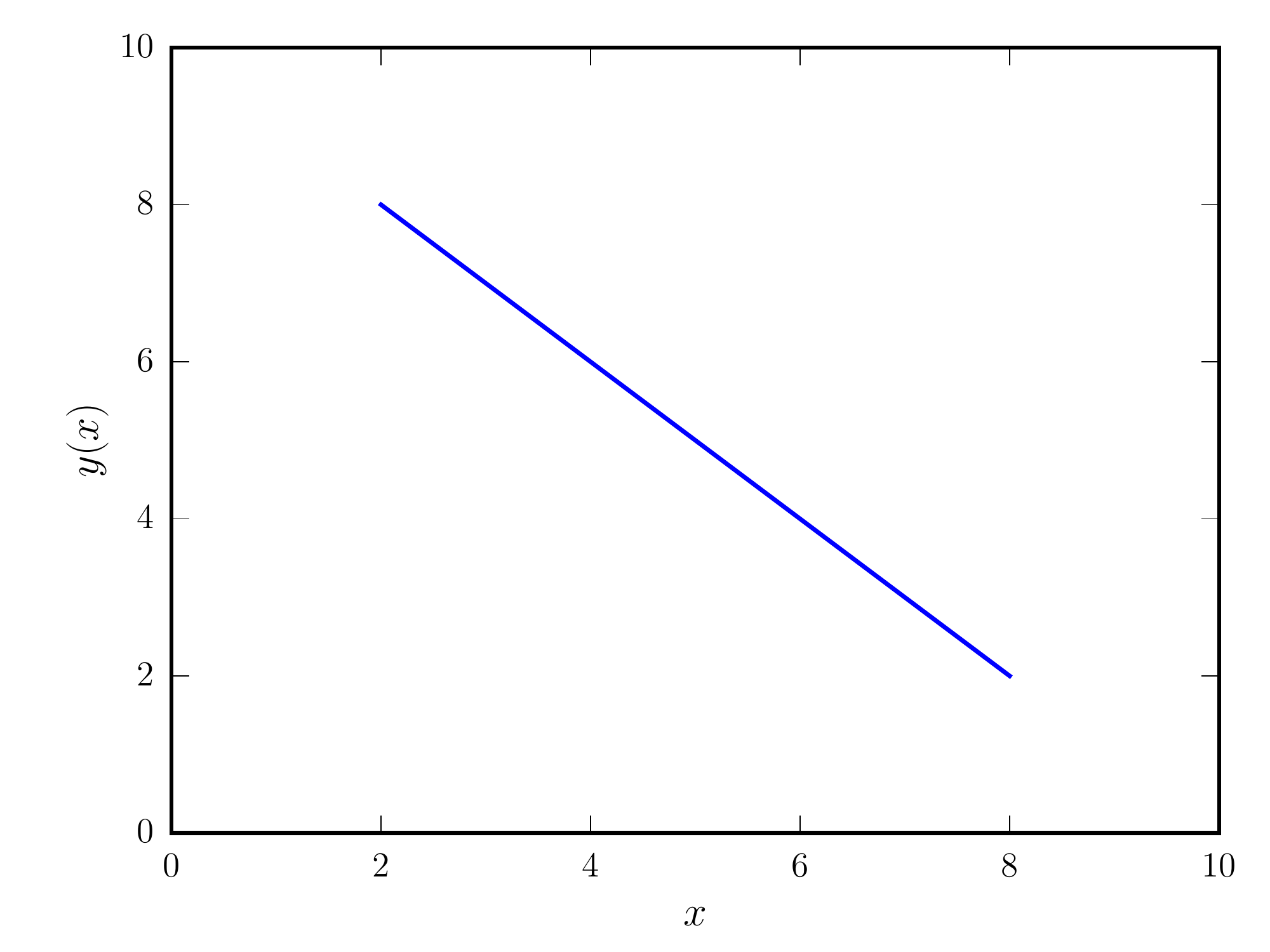}\\
\includegraphics[width= \linewidth]{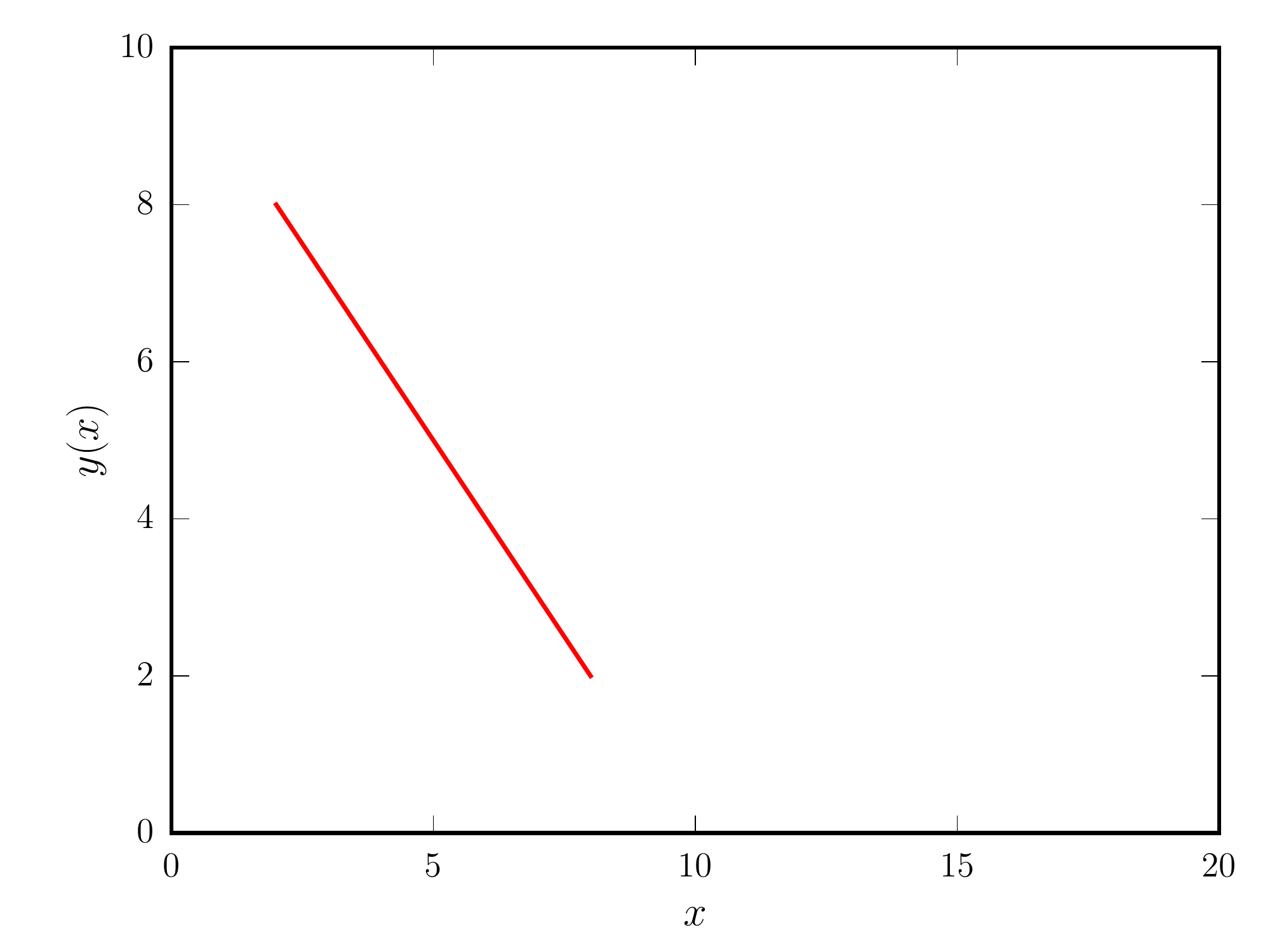}
\caption{Comparing $y$ vs $x$ plots due to the squeezing of $x$ axis.}
\label{Fig:shrinkxsq}
\end{figure}

These phenomena can easily be interpreted by the fact that `e-fold' clocks $\Delta$ differ in different frames. As discussed before, equation (\ref{Eq:EquivRelAttrac}) tells us that if $\Delta N$ becomes smaller in one frame than the other, $|\lambda|$ in the corresponding frame becomes higher than the other. This is perfectly demonstrated in Figure \ref{Fig:shrinkxsq}. In this Figure, we plot $y$ vs $x$ on the left with the slope of unity and the same graph by squeezing the $x$-axis by half on the right. As a result, the slope in the later looks steeper than the earlier but two units of $x$ in the first one are equivalent to one unit of $x$ in the second one. Mathematically, if $m$ is the slope of the first plot and $\tilde{m}$ is the slope of the next one, then since $y$ axes remain unchanged in these two plots, $$m\,\Delta x = \tilde{m}\,\Delta \tilde{x}\Rightarrow \tilde{m} = m\,\frac{\Delta x}{\Delta\tilde{x}}.$$Since, $m = 1$ and $\Delta x/\Delta\tilde{x} = 2$ due to the squeezing, $\tilde{m}$ becomes $2$, and hence, looks steeper than that of before. The Brans-Dicke theory and the corresponding Einstein frames are analogous to the first and second plots, respectively. In the Einstein frame, one e-fold is equivalent to a higher number of e-folds in the Brans-Dicke theory and as a result, deviation decays faster in Einstein frame than the same in the Brans-Dicke theory due to the equivalence (\ref{Eq:EquivRelAttrac}).

The result is \emph{interesting} as well as \emph{compelling}. Since the background quantities are related by the conformal relations, it is expected that the deviations in different conformally connected frames are proportional. This can be shown easily as $$ \tilde{x} = f(x) \quad \Rightarrow \quad \delta \tilde{x} = f'(x)|_*\,\delta x$$ and therefore, we expect the stability to be identical. However, even though the deviations are proportional to each other, the difference in stability conditions arises due to the fact that the e-folds differ from one frame to other.

To illustrate the above phenomena in the context of slow-roll inflation, we will carefully study different inflationary models in two different frames and see the differences in a more pragmatic manner in the next section.

\section{Results for inflationary models}\label{Sec:InflationaryContext}

In this section, we focus on the inflationary scenario and consider attractor models which give rise to slow-roll inflation in both frames. In doing so, we examine two inflationary models: Starobinsky \cite{STAROBINSKY198099, STAROBINSKY1982175} and Chaotic inflation \cite{Linde:1983gd}. Starobinsky model had been proposed as an extension of Einstein gravity by adding higher order curvature corrections, i.e., it falls under $f(R)$ gravity theory. The action for general $f(R)$ theory is given by

\begin{eqnarray}\label{Eq:Actionf(R)}
\mathcal{S}_{f(R)} = \frac{M_{\rm pl}^2}{2}\,\int {\rm d}^4 x\,\sqrt{-g}\,f(R),
\end{eqnarray} 
which can again be re-written by introducing auxiliary fields as
\begin{eqnarray}\label{Eq:Actionf(R)Jordan}
&&\mathcal{S}_{f(R)} = \frac{M_{\rm pl}^2}{2}\,\int {\rm d}^4 x\,\sqrt{-g}\,\left[f(S) + \varphi\left(R - S\right)\right] \nonumber\\
&&\qquad= \frac{M_{\rm pl}^2}{2}\,\int {\rm d}^4 x\,\sqrt{-g}\,\left[ \varphi R - 2\frac{\left(\varphi S(\varphi) - 2f(S(\varphi))\right)}{2} \right].\nonumber\\
\end{eqnarray}
\noindent $S$ and $\varphi$ are not independent but related by the equation $\varphi = f'(S)$. The field $\varphi$ is referred to as `scalaron'. Therefore, any $f(R)$ theory is same as Brans-Dicke theory (\ref{Eq:ActionBDGeneral}) with Brans-Dicke parameter $\omega_{\rm BD} = 0$ and the potential is given by

\begin{eqnarray}\label{Eq:Potentialf(R)Jordan}
V(\varphi) = M_{\rm pl}^2 \,\frac{\left(\varphi S(\varphi) - 2f(S(\varphi))\right)}{2}.
\end{eqnarray} 

\noindent Notice the change of $M_{\rm pl}^2$ factor in the potential as the factor has already been introduced in (\ref{Eq:Actionf(R)}). Hence, in case of $f(R)$ theory, the redefined scalar field and the potential in conformal Einstein frame take the form
\begin{eqnarray}\label{Eq:Potentialf(R)Einstein}
\varphi = \exp\left(\sqrt{\frac{2}{3}}\frac{\tilde{\varphi}}{M_{\rm pl}}\right), \tilde{V}(\tilde{\varphi}) =  \frac{V(\varphi(\tilde{\varphi}))}{\varphi(\tilde{\varphi})^2}.
\end{eqnarray}

\noindent Therefore, by using the reverse transformation of the above relations, any theory can be conformally mapped to $f(R)$ theory. We will use this in case of chaotic inflation.

The reason of considering equivalent $f(R)$ model and not any other Jordan frame theory is as follows.  In case of slow-roll inflation, the scale factors in $f(R)$ theory and its equivalent Einstein frame behave nearly similarly. This can easily be shown by using the relation (\ref{Eq:wbdq}). In case of $f(R)$ theory, as we have already shown, $\omega_{\rm BD} = 0$ and therefore
\begin{eqnarray}
	&&\omega_{\rm BD} = 0 \quad \Rightarrow \quad(n - \alpha)^2 = \frac{1}{3}\, \alpha \,(1 + \alpha) \nonumber\\
	 &&\Rightarrow \quad n \approx \alpha \quad \mbox{for}\quad \alpha \approx -1.\nonumber
\end{eqnarray}
Therefore, if the scale factor in the Einstein frame is nearly de-Sitter, the scale factor in the corresponding $f(R)$ theory is also nearly de-Sitter.

\subsection{Starobinsky inflation}\label{Sec:Starobinsky}

The action for the Starobinsky model \cite{STAROBINSKY1982175} is

\begin{eqnarray}\label{Eq:Starof(R)}
\mathcal{S}_{f(R)} =  \frac{M_{\rm pl}^2}{2}\,\int {\rm d}^4 x\,\sqrt{-g}\,\left(R + \frac{1}{6 m^2}\,R^2\right).
\end{eqnarray}

\noindent $m$ is the model parameter and from the observation, it can be fixed as $m \approx 10^{-5} \,M_{\rm pl}$. Using (\ref{Eq:Actionf(R)Jordan}) and (\ref{Eq:Potentialf(R)Jordan}), we can find the potential in the Jordan $f(R)$ frame for the Starobinsky model as

\begin{eqnarray}
V(\varphi) = \frac{3}{4}\, m^2\, M_{\rm pl}^2\, (1 - \varphi)^2.
\end{eqnarray} 

\noindent The corresponding potential in the Einstein frame can easily be obtained by using the relation (\ref{Eq:Potentialf(R)Einstein}) as

\begin{eqnarray}
\tilde{V}(\tilde{\varphi}) = \frac{3}{4}\, m^2\,M_{\rm pl}^2\,\left(1 - \exp\left[-\sqrt{\frac{2}{3}}\frac{\tilde{\varphi}}{M_{\rm pl}}\right]\right)^2
\end{eqnarray}

\begin{figure}
\includegraphics[width = \linewidth]{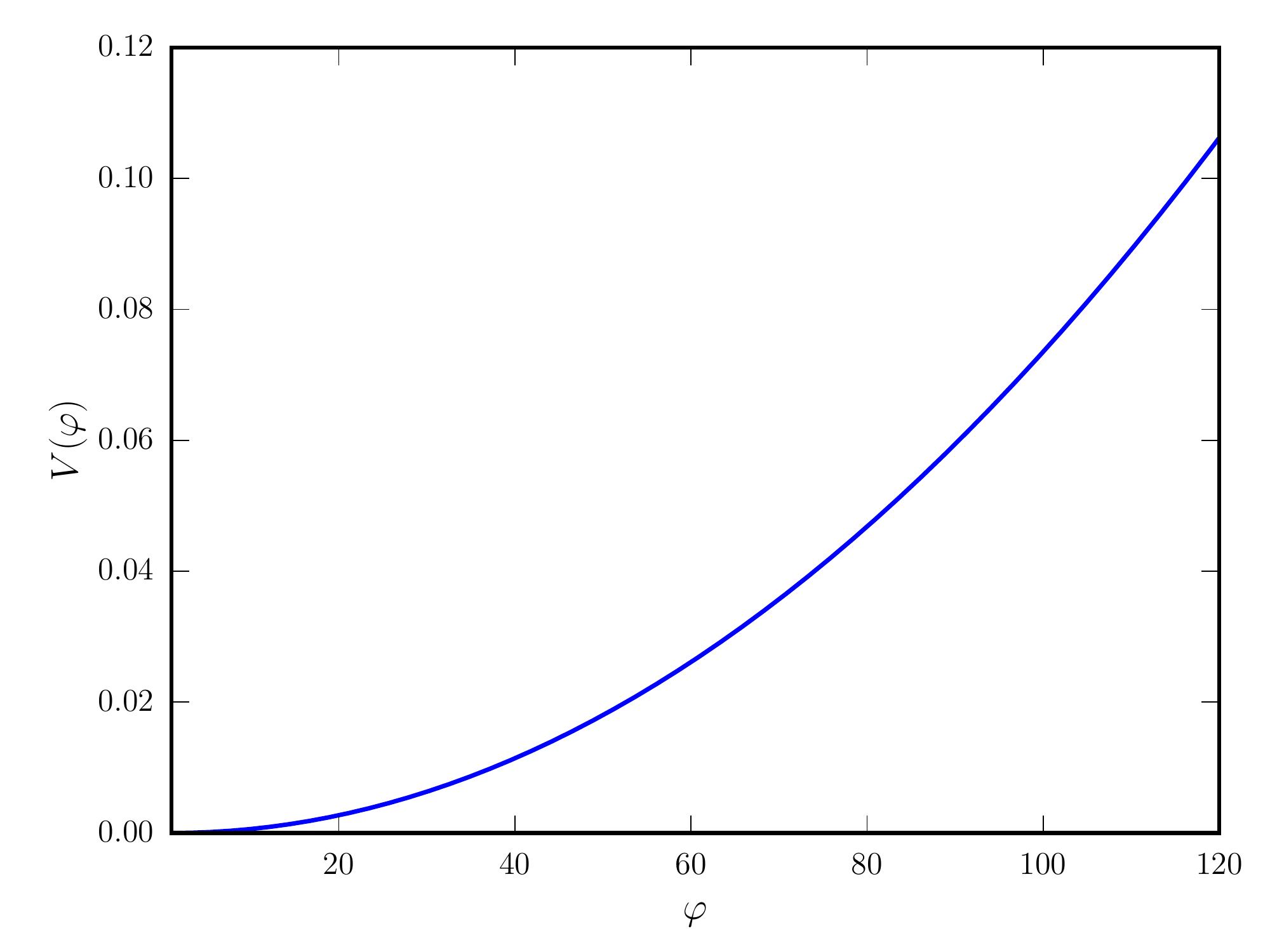}\\
\includegraphics[width = \linewidth]{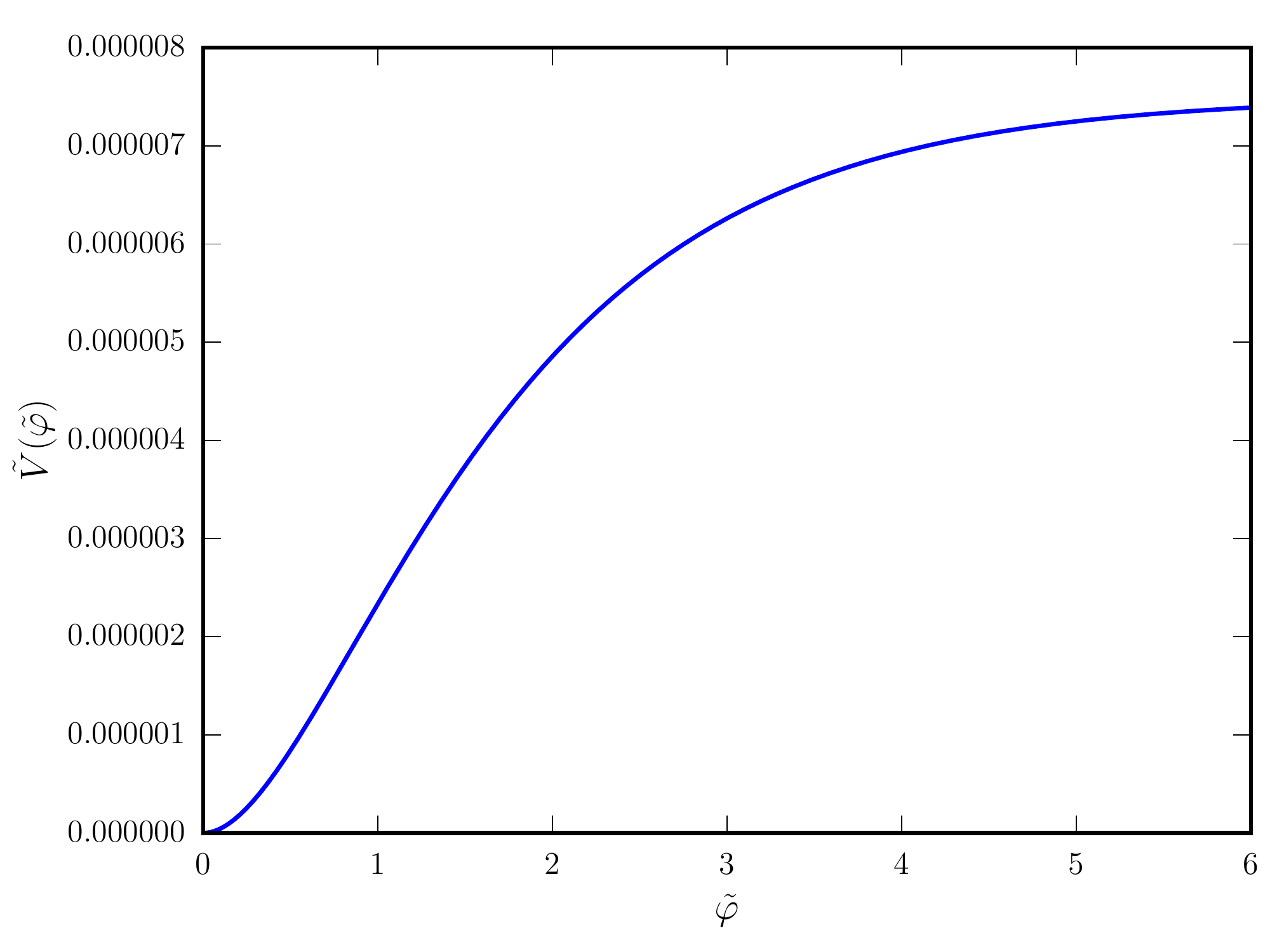}
\caption{Starobinsky potential in the Jordan $f(R)$ frame (top) and in the Einstein frame (bottom). Planck mass ($M_{\rm Pl}$) is set to unity.}	
\label{Fig:STPot}
\end{figure}

\noindent Both the potentials are shown in Figure \ref{Fig:STPot}.

Equations (\ref{Eq:DeltaN}) and (\ref{Eq:LambdaFrames}) imply that, for an attractor solution, if $ \,\tilde{\varphi}_N/M_{\rm pl} > 0$   is positive, then $\lambda > \tilde{\lambda}$, i.e., in the Jordan $f(R)$ frame, deviation decays faster than the same in the Einstein frame and for $\tilde{\varphi}_N/M_{\rm pl} < 0$, the opposite happens. For single field slow-roll inflation in the Einstein frame, $\tilde{\varphi}_N/M_{\rm pl} \lesssim 0$. Therefore, deviation in Einstein frame decays faster than the same in the Jordan $f(R)$ frame. In order to show this, we consider equivalent initial conditions in both frames:

\begin{eqnarray}
\varphi &=& \exp \left[\sqrt{\frac{2}{3}} \tilde{\varphi}\right] \\
\varphi_N &=& \sqrt{\frac{2}{3}}\, \frac{\exp \left[\sqrt{\frac{2}{3}} \tilde{\varphi}\right]}{1 -  \frac{1}{\sqrt{6}}\, \tilde{\varphi}_{\tilde{N}}}\, \tilde{\varphi}_{\tilde{N}}
\end{eqnarray}
where $\varphi_N \equiv \partial\varphi/\partial N,\, \tilde{\varphi}_{\tilde{N}} \equiv \partial\tilde{\varphi}/\partial \tilde{N}$. Therefore, once we fix the initial conditions, i.e, $(\varphi, \, \varphi_N)$ in one frame, the initial conditions in other frame are automatically fixed. Once the initial conditions are fixed, using equations (\ref{Eq:0-0BackBD}), (\ref{Eq:AccBackBD}) and (\ref{Eq:ScBackBD}) we can numerically solve the Starobinsky model in the Jordan $f(R)$ frame. Similarly, we can solve the model in the Einstein frame by using the equations (\ref{Eq:EoMsBackEinst1}), (\ref{Eq:EoMsBackEinst2}) and (\ref{Eq:EoMsBackEinst3}).

In order to show the differences in attractor behavior in these two frames, we consider two different sets of initial conditions which represent significant deviations from the slow-roll fixed point $\tilde{\varphi}_{\tilde{N}} \lesssim 0$ and have $70-90$ e-folding of inflation. These are:
\begin{itemize}
	\item[1.] {\bf Field velocity is highly negative:}  $ \tilde{\varphi}_{\tilde{N}}(0) = -\sqrt{5.99}$.
	\item[2.] {\bf Field velocity is highly positive:}  $\tilde{\varphi}_{\tilde{N}}(0) = 1$.
\end{itemize}

\begin{figure}
	\includegraphics[width = \linewidth]{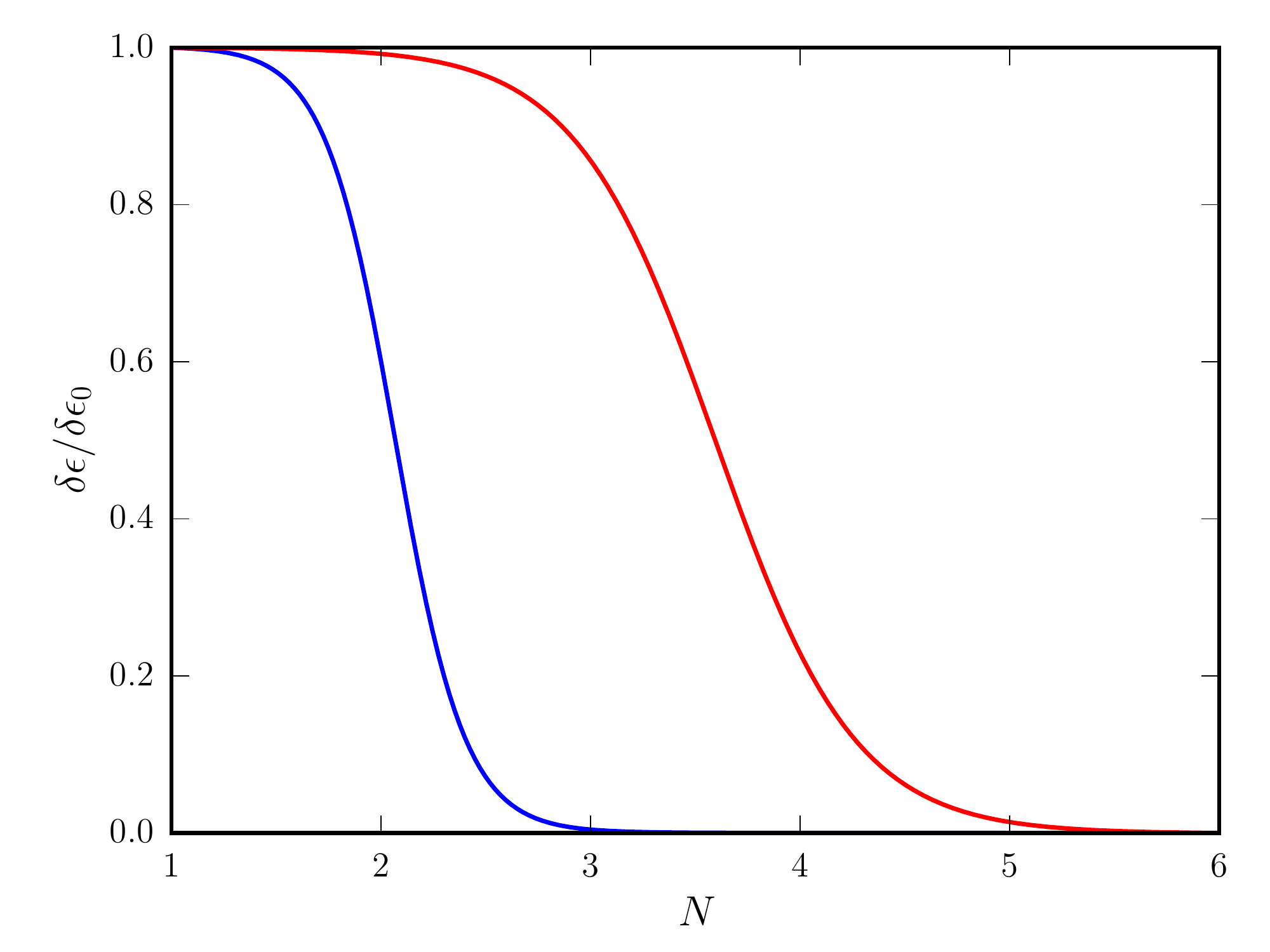}\\
	\includegraphics[width = \linewidth]{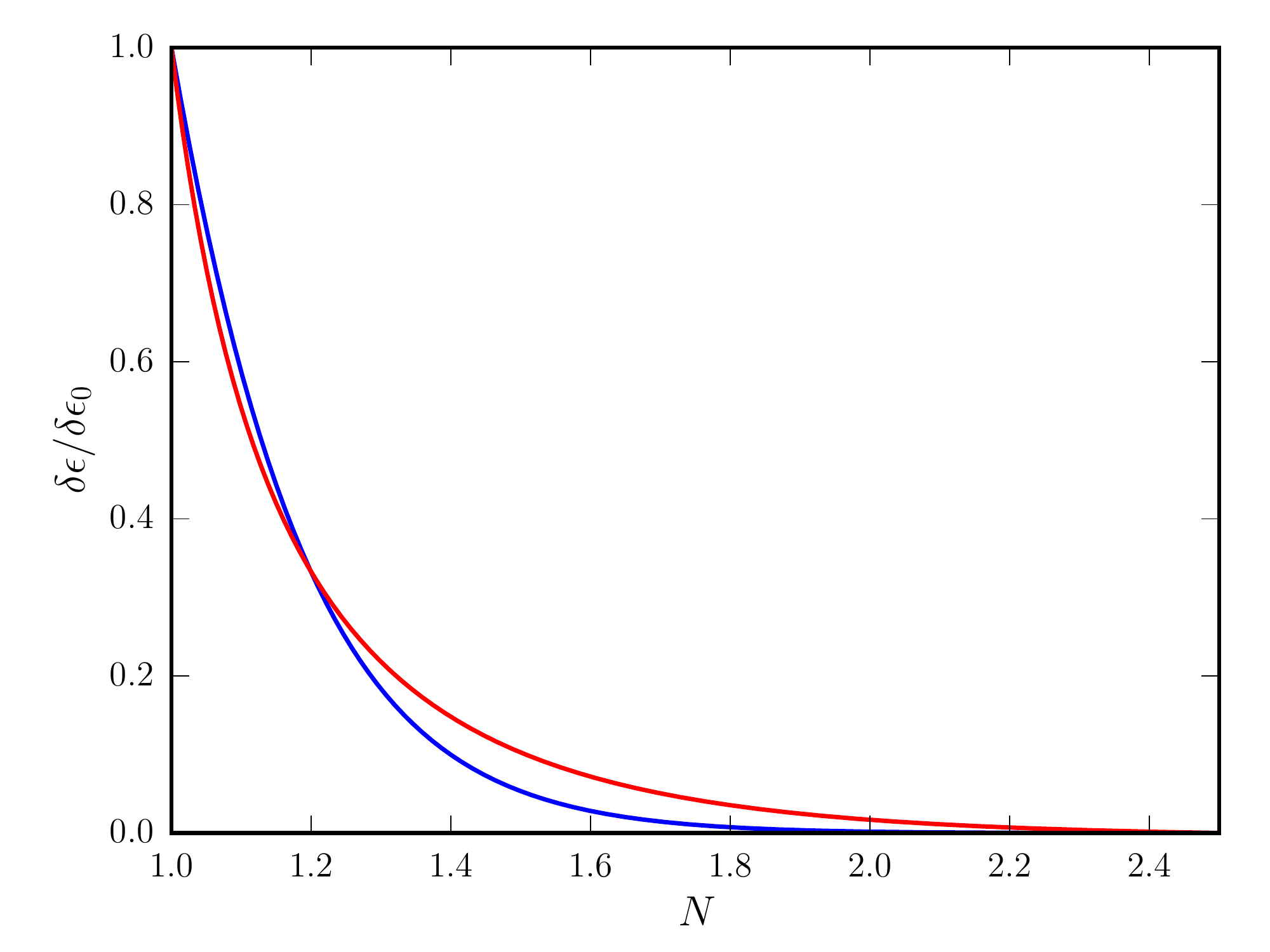}
	\caption{Deviations of the slow-roll parameters in both the Einstein frame (color blue) and the Jordan $f(R)$ frame (color red) are plotted for highly negative field velocity (top) and highly positive field velocity (bottom) in the Einstein frame. Deviations are properly normalized.}
	\label{Fig:devSt}
\end{figure}

\begin{figure}
\includegraphics[width = \linewidth]{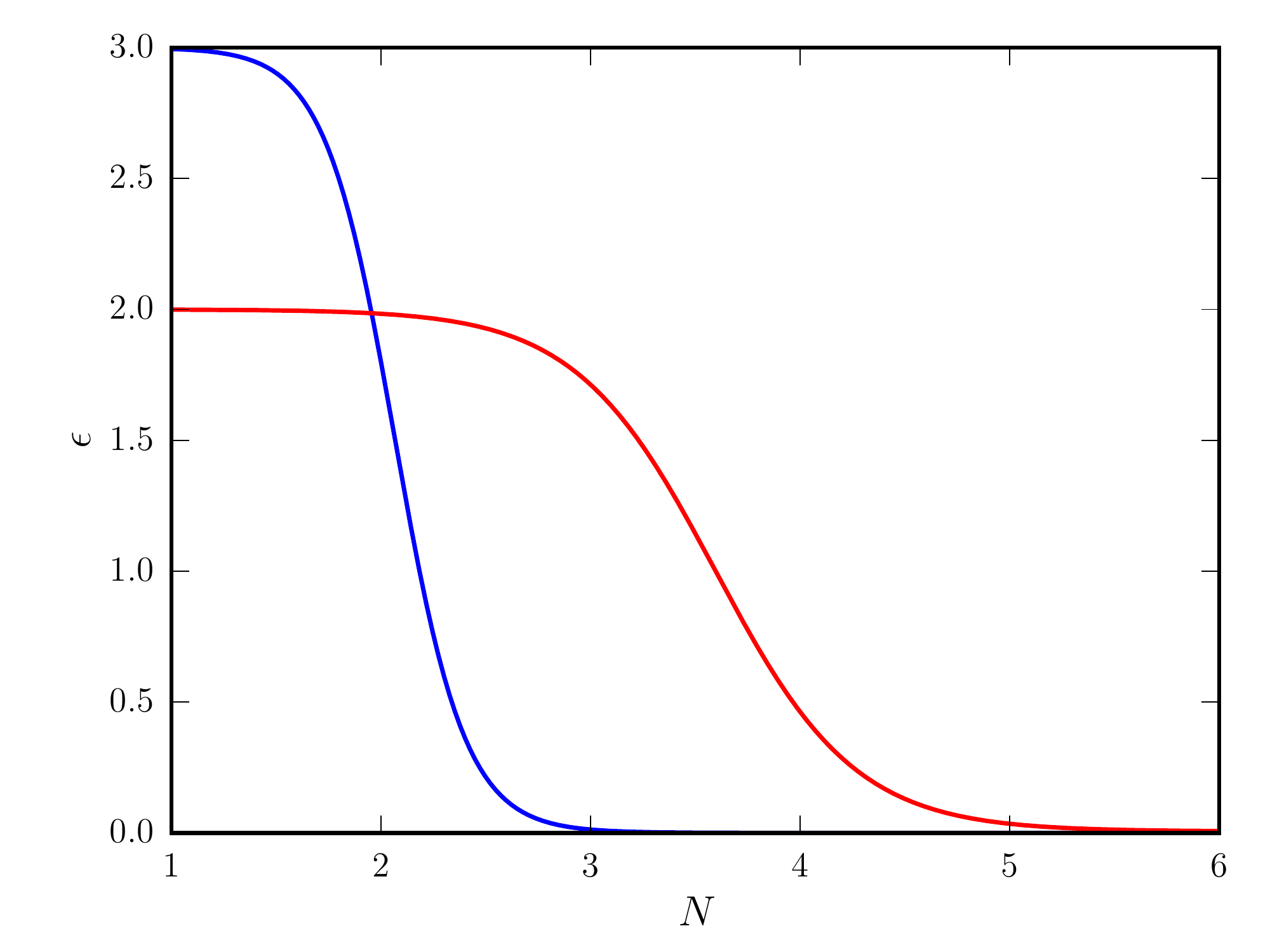}\\
\includegraphics[width = \linewidth]{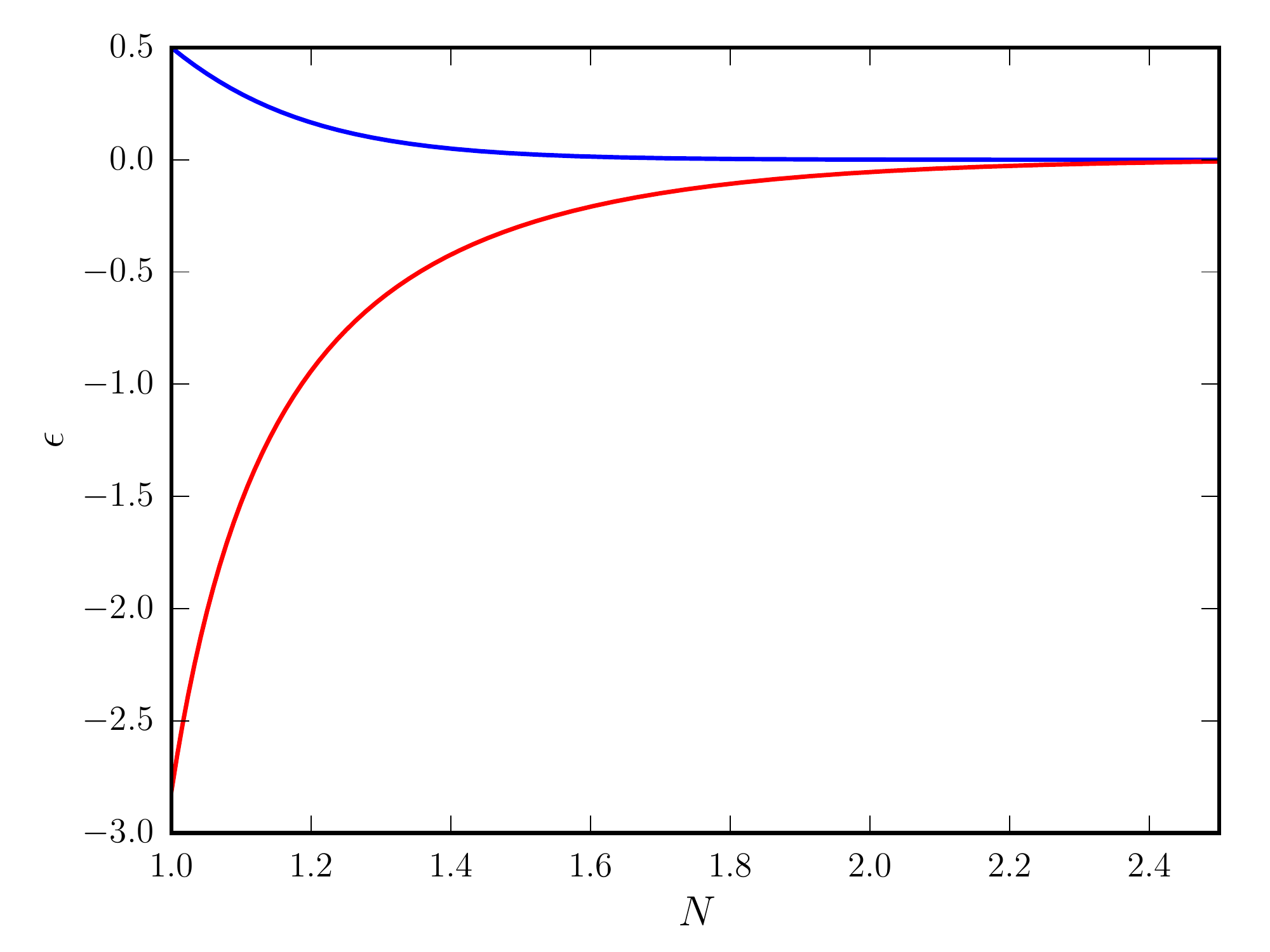}
\caption{Slow-roll parameters in both the Einstein frame (color blue) and the Jordan $f(R)$ frame (color red) are plotted for highly negative field velocity (top) and highly positive field velocity (bottom) in the Einstein frame.}
\label{Fig.devSt'}
\end{figure}

The results are plotted in Figures \ref{Fig:devSt} and \ref{Fig.devSt'} ($M_{\rm pl}$ is taken to be unity). The blue line is the numerical solution in the Einstein frame and the red line represents the same in the Jordan $f(R)$ frame. In the first case where field velocity is highly negative, we take $\tilde{\varphi}(0)$ to be $9$ whereas in the next case, $\tilde{\varphi}(0)$ is fixed at $5.5$. However, it can be fixed at any value. 

Notice the difference in initial conditions in these two cases. In the first case where the field velocity in the Einstein frame is highly negative,  the initial value of the slow-roll parameter, $\tilde{\epsilon}(0)$ in the Einstein frame is $3.0$, whereas,  in the Jordan $f(R)$ frame, it is $2.0$ (at the top in Figure \ref{Fig.devSt'}). This implies that the initial deviation in the Einstein frame is higher than that in the Jordan $f(R)$ frame. In the second case where the field velocity in the Einstein frame is positive, while the initial value of the slow-roll parameter in the Einstein frame is $0.5$, in the Jordan $f(R)$ frame, it possesses a highly negative value of $-3.0$ (super-inflation) (at the bottom in Figure \ref{Fig.devSt'}). We infer the following from Figures \ref{Fig:devSt}:

\begin{itemize}
	\item[1.] {\bf Field velocity is highly negative:} The deviation in the Einstein frame takes roughly 3 e-folds to decay, whereas, it takes almost 5.5 e-folds in the Jordan $f(R)$ frame to reach the slow-roll state.
	\item[2.] {\bf Field velocity is highly positive:} The deviation in the Einstein frame takes roughly 1.8 e-folds to decay, whereas, it takes almost 2.4 e-folds in the Jordan $f(R)$ frame to reach the slow-roll state.
\end{itemize}

Both cases imply that irrespective of the differences in the initial conditions, the deviations decay faster in the Einstein frame than in the Jordan $f(R)$ frame.

\begin{figure}[H]
	\includegraphics[width = \linewidth]{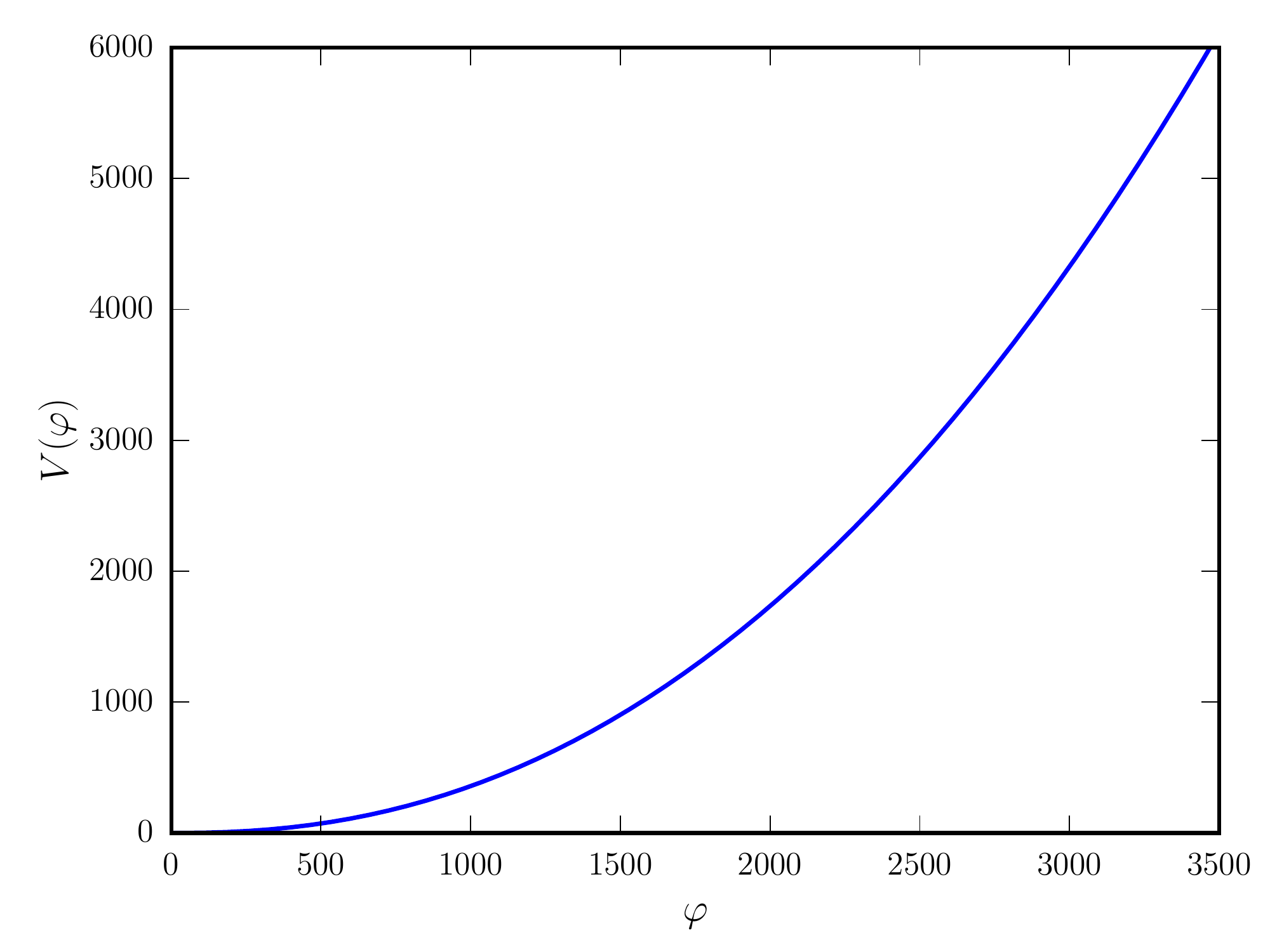}\\
	\includegraphics[width = \linewidth]{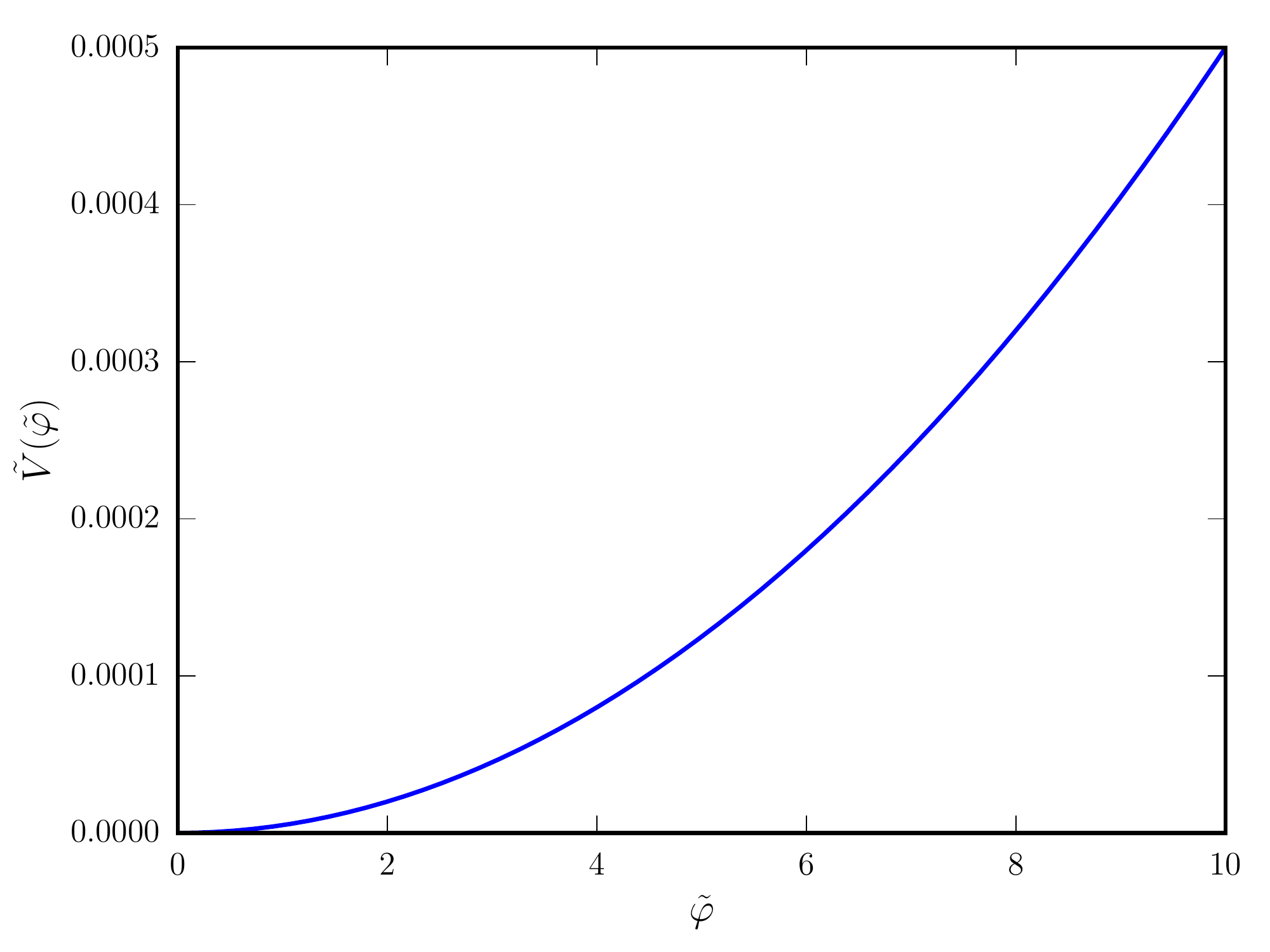}
	\caption{Chaotic potential in both the Jordan $f(R)$ frame (top) and the Einstein frame (bottom). Planck mass ($M_{\rm Pl}$) is set to unity.}
	\label{Fig:CPot}
\end{figure}

\subsection{Chaotic inflation}\label{Sec:Chaotic}

In the case of chaotic inflation \cite{Linde:1983gd}, the scalar field potential in the Einstein frame is

\begin{eqnarray}
\tilde{V}(\tilde{\varphi}) = \frac{1}{2}\,m^2 \,\tilde{\varphi}^2.
\end{eqnarray}

\noindent In order to find the corresponding redefined scalar field and the potential in the Jordan $f(R)$ frame, we can use the inverse of the relation (\ref{Eq:Potentialf(R)Einstein}) and these are

\begin{eqnarray}
\tilde{\varphi} = \sqrt{\frac{3}{2}} M_{\rm pl}\, \ln\,\varphi, \quad V(\varphi) = \frac{3}{4}\,m^2 M_{\rm pl}^2 \,\left(\varphi\, \ln\,\varphi\right)^2.
\end{eqnarray}

\begin{figure}[H]
	\includegraphics[width = \linewidth]{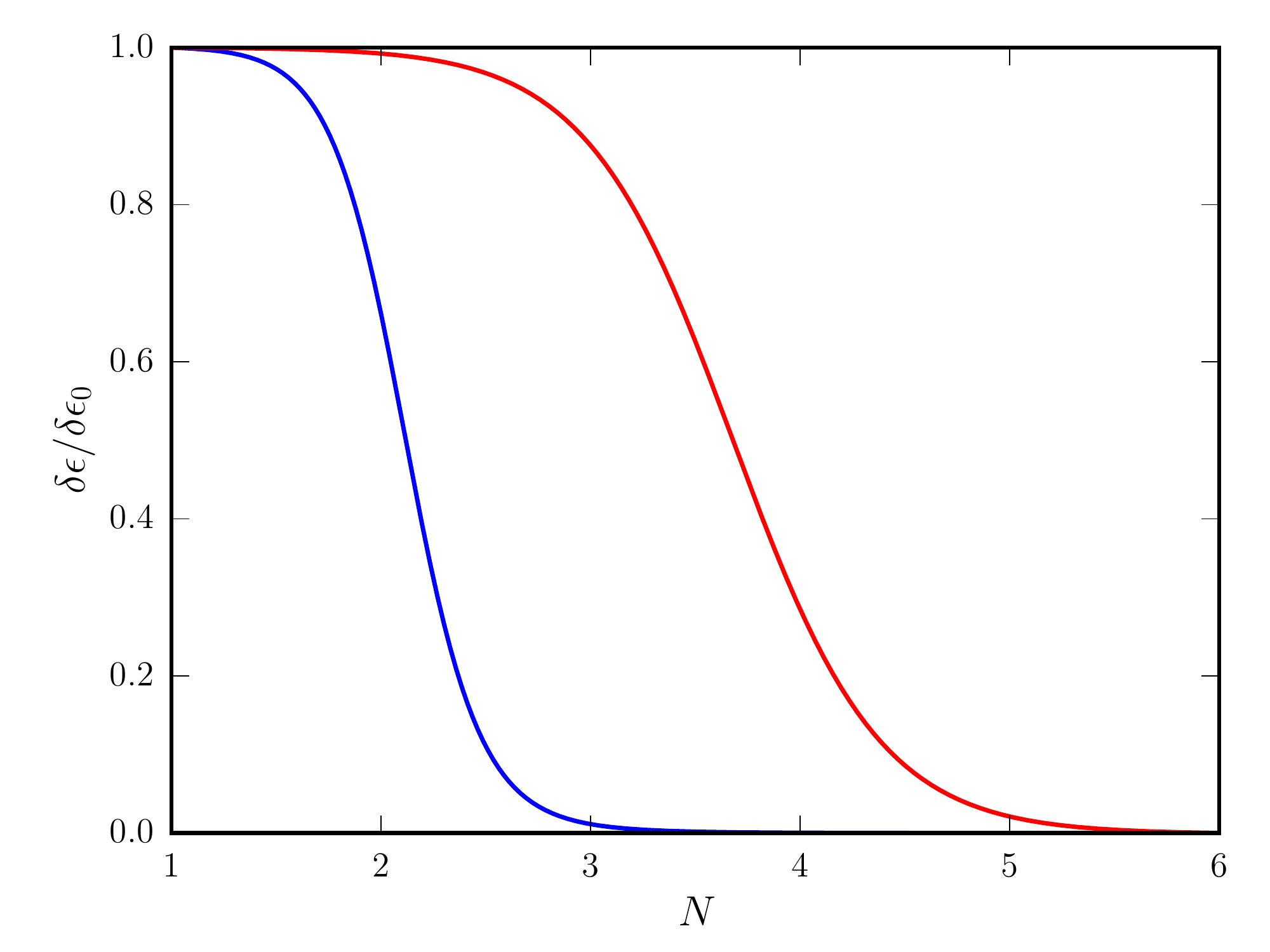}\\
	\includegraphics[width = \linewidth]{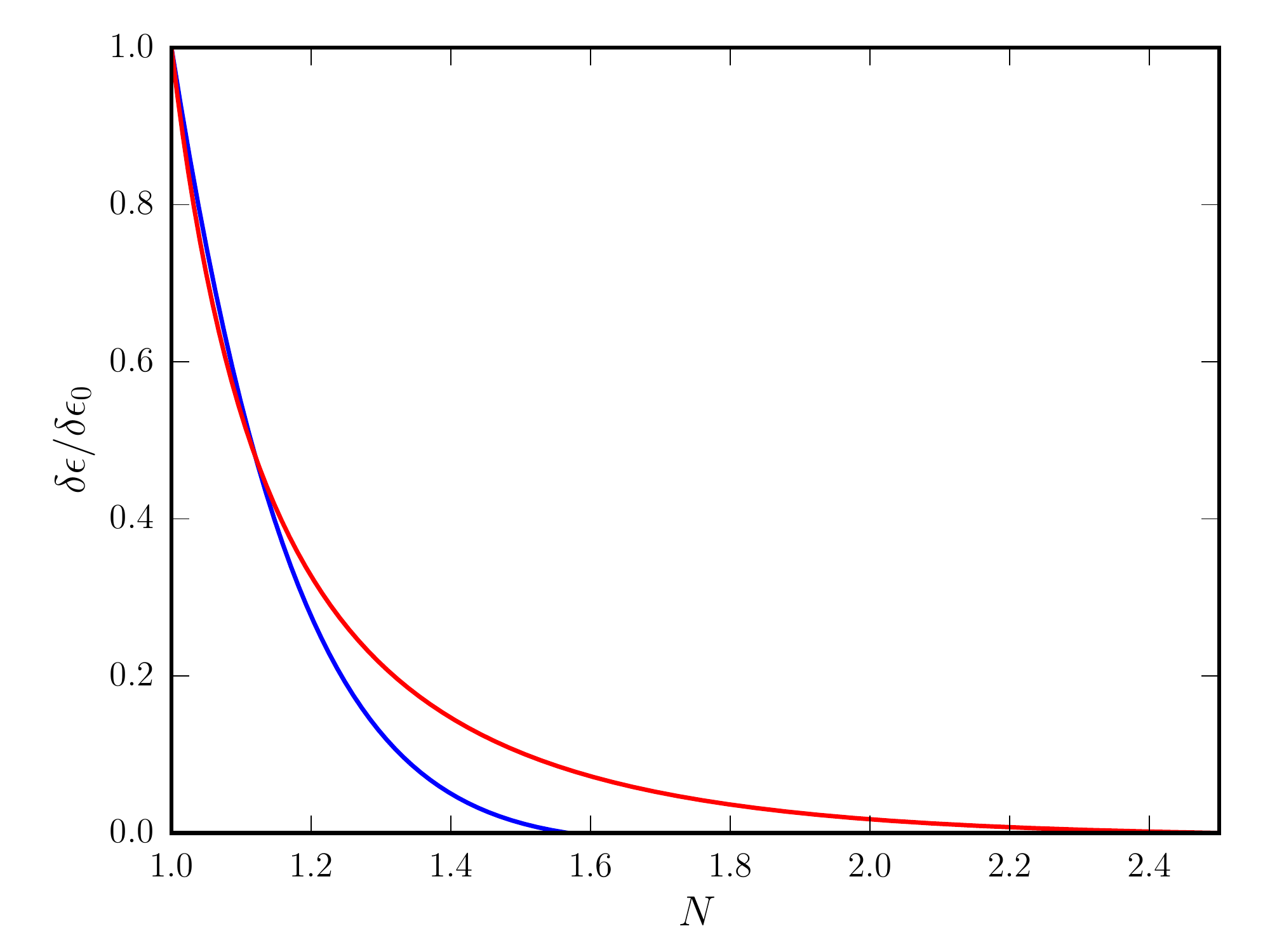}
	\caption{Deviations of the slow-roll parameters in both the Einstein frame (color blue) and the Jordan $f(R)$ frame (color red) are plotted for highly negative field velocity (top) and highly positive field velocity (bottom) in the Einstein frame. Deviations are properly normalized.}
	\label{Fig:SLPC}
\end{figure}

\begin{figure}
	\includegraphics[width = \linewidth]{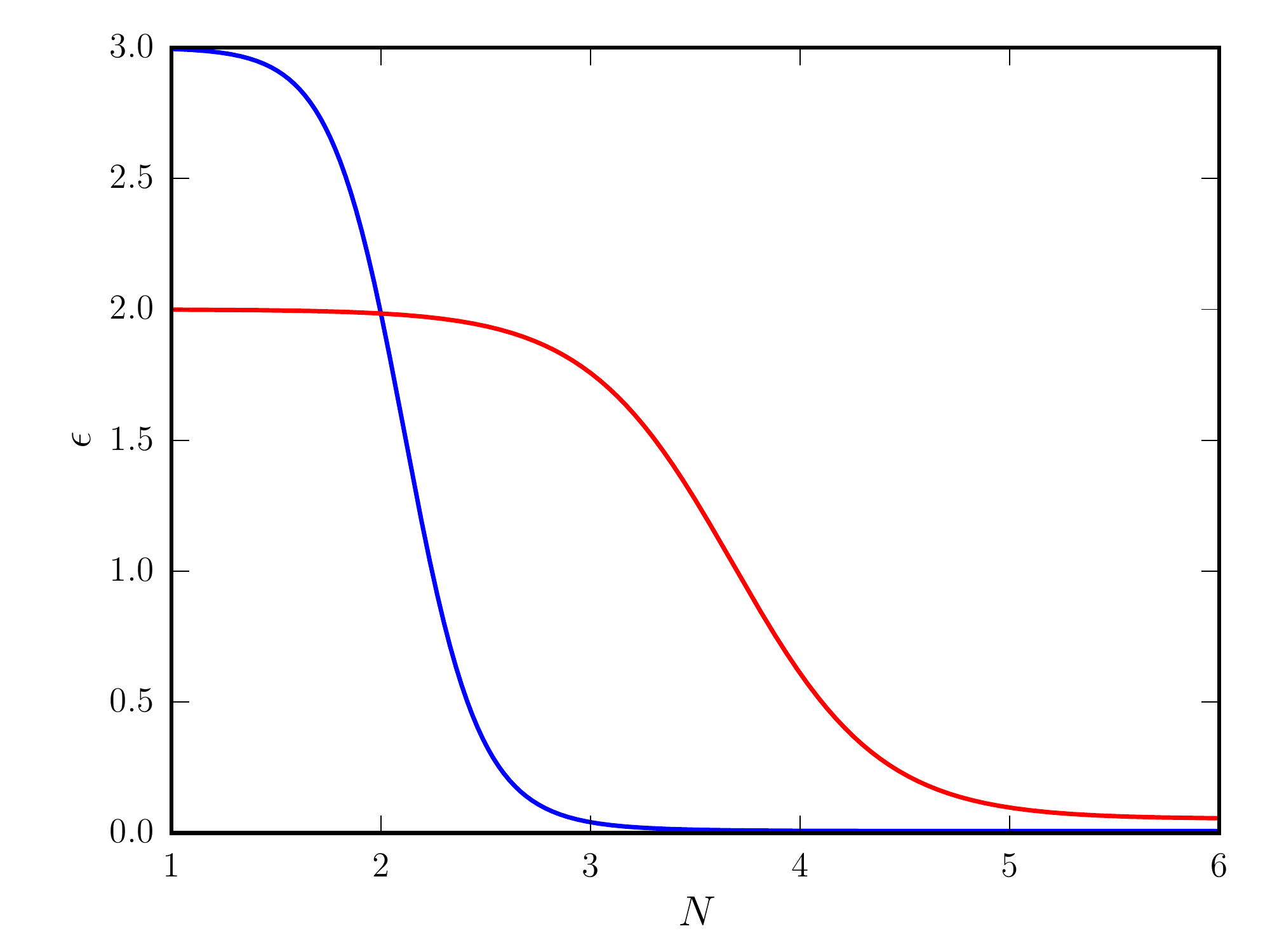}\\
	\includegraphics[width = \linewidth]{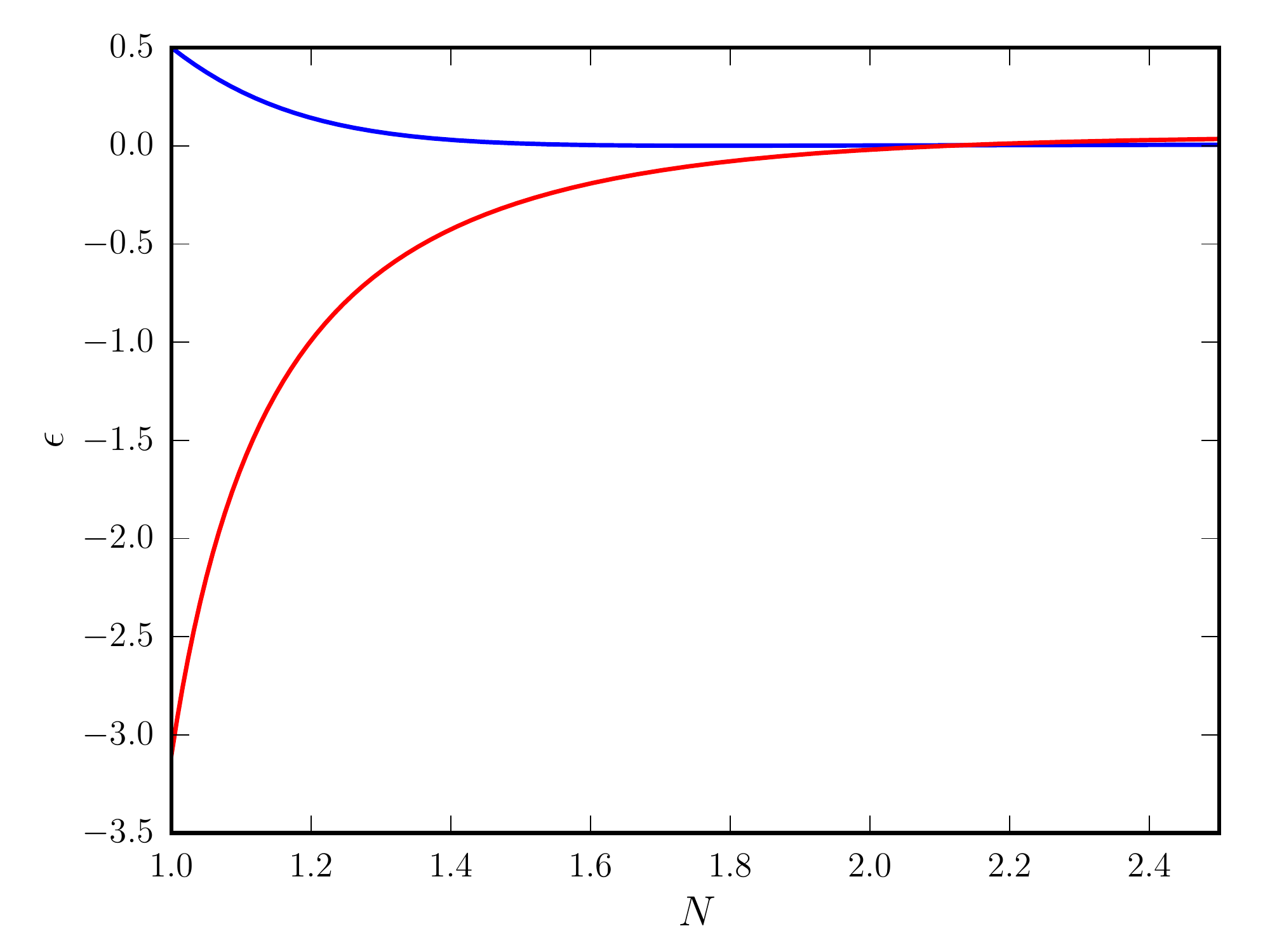}
	\caption{Slow-roll parameters in both the Einstein frame (color blue) and the Jordan $f(R)$ frame (color red) are plotted for highly negative field velocity (top) and highly positive field velocity (bottom) in the Einstein frame.}
	\label{Fig:SLPC'}
\end{figure}

\noindent Both the potentials are shown in Figure \ref{Fig:CPot} ($M_{\rm pl}$ is taken to be unity). Again we consider two extreme deviations from the slow-roll state, i.e., two different domain of initial conditions which are identical to Starobinsky inflation and the inflationary phase lasts for $70 - 90$ e-folds. In the first case where field velocity is highly negative, we take $\tilde{\varphi}(0)$ to be equal to $20$ and in the next case, $\tilde{\varphi}(0)$ is fixed at $17$. The solutions are shown in Figures \ref{Fig:SLPC} and \ref{Fig:SLPC'}. The result repeats the same as the Starobinsky model:

\begin{itemize}
	\item[1.] {\bf Field velocity is highly negative:} The deviation in the Einstein frame takes roughly 3 e-folds to decay, whereas, it takes almost 5.5 e-folds in the Jordan $f(R)$ frame to reach the slow-roll state.
	\item[2.] {\bf Field velocity is highly positive:} The deviation in the Einstein frame takes roughly 1.8 e-folds to decay, whereas, it takes almost 2.4 e-folds in the Jordan $f(R)$ frame to reach the slow-roll state.
\end{itemize}

Therefore, also, in this case, the deviation in the Einstein frame decays faster than the same in the Jordan $f(R)$ frame. The result can be generalized for any attractor slow-roll models of inflation.


\section{Difference in e-folds during inflation}\label{Sec:e-folds}

Again, by looking at the relation (\ref{Eq:DeltaN}), one can easily notice that in the Einstein frame, the `e-fold' interval $\delta N$ is smaller than that in the Jordan frame during slow-roll inflation. Therefore, for a given equivalent interval, in the Jordan frame, the total number of e-folds should be higher than the same in the Einstein frame. Again, this is perfectly illustrated in Figure \ref{Fig:shrinkxsq}. In this Figure, $\Delta \tilde{x} = 1/2\,\Delta x$, and hence, $10$ intervals of $x$ is equivalent to $5$ intervals of $\tilde{x}$. Similarly, since  $\tilde{\varphi}_{\tilde{N}}$ is negative but close to zero during the slow-roll inflationary period, in the Einstein frame, the total number of e-folds is lesser than the same in the Jordan frame for the equivalent duration of an event. We can use this to evaluate the duration of inflation for both Starobinsky as well as chaotic inflation in both these frames. In order to show the difference of e-folds during inflation in these two frames, we need to use the definition of end of inflation at $N = N_{\rm end}$ and it is $\epsilon(N_{\rm end}) = 1$, i.e., the slow-roll parameter becomes unity at the end of inflation.

In Table \ref{Tab:Diff}, we provide the results for both the models in these two frames. We use the same initial conditions that we used before. As you can see, in both Starobinsky and chaotic inflation, the number of e-folds in the Jordan frame is higher than the same in the Einstein frame.

\section{Conclusions}

In this work, we studied the effect of conformal transformation on the attractors in the context of the early Universe. We specifically studied the stability of the fixed points in the Brans-Dicke theory and in the corresponding conformal Einstein frame, and explicitly showed the differences. In doing so, we first considered the power law cosmology in both the frames and carefully studied the phase space behavior without the additional fluid.

Since all the equations are equivalent in conformal frames, one expects equivalent attractor behavior of the solutions in these two frames.  We found that they are indeed equivalent, i.e., attractor behavior in one frame guarantees conformal solutions in other frames to also be attractors. However,  attractor behaviors in both frames differ depending upon the dynamics of the frame. In the case of power law cosmology, the difference depends on the exponents of the scale factors in both frames. We also showed that the e-fold time interval changes from frame to frame and we established the equivalence relation of the stability in all conformally connected frames: $\lambda\, \Delta N$ to be invariant under all frames (see relation (\ref{Eq:EquivRelAttrac})). If $\Delta N$ in one frame shrinks more than that in the other frame, in order to maintain the equivalence, the corresponding $\lambda$, which governs the dynamics of the deviation, becomes higher than the same in another frame. This implies that deviations in that frame decay faster than the same in the other frame.

The result can easily be extended to any generalized model in any conformally connected frames. The relation (\ref{Eq:EquivRelAttrac}) is indeed the general equivalence relation of the attractor behavior in any conformally connected frames. In order to study the behavior in more realistic models, we consider two popular inflationary models: Starobinsky and Chaotic inflation in the Einstein frame as well as in the Jordan $f(R)$ frame. In slow-roll inflation in the Einstein frame, $\varphi_N \lesssim 0$ and therefore in the Einstein frame, the e-fold time interval is smaller  than that in the corresponding $f(R)$ theory (see relation (\ref{Eq:DeltaN})) and therefore, $\lambda$, the decaying parameter for an attractor solution is higher in the Einstein frame compared to the other. Therefore, in the case of slow-roll inflation, deviations decay faster in the Einstein frame than that in the Jordan $f(R)$ frame. On the other hand, since the time shrinks in the Einstein frame more than that in the Jordan $f(R)$ frame, we evaluate the duration of inflation in both the frames and we found that inflation lasts longer in the Jordan $f(R)$ frame than in the Einstein frame.

Since we found that the time axis squeezes or expands, it can now be easily be related to the definition of conformal transformation itself. Under conformal transformation, the space-time in different frames scales differently, which can be seen from the relation (\ref{Eq:ConfTrans}). Therefore, an event in different conformal frames appears to be squeezed or stretched depending upon the conformal behavior in the respective frames.   

There are two physical implications of the above phenomena. First, following the inflationary phase, reheating generally arises due to an additional relativistic fluid(s). Since the attracting behavior, as well as the number of e-folds during inflation, differ in these two frames, we may see a significant signature of the phenomena in the reheating phase. Second, attractor solution can produce features in the perturbed power spectrum \cite{Sreenath:2014nca} which help to improve the fit to the observation. Since the behavior is different in different frames, the features in these two frames will be different. These works along with other observational consequences are under progress.

\section*{Acknowledgment}     
I thank L. Sriramkumar and S. Shankaranarayanan for useful and important discussions as well as their valuable comments on this work.  I also  wish to
thank the Indian Institute of Technology Madras, Chennai, India, for support through the Exploratory Research Project PHY/17-18/874/RFER/LSRI.

\onecolumngrid

\begin{table}[!b]
	\begin{center}
		\begin{tabular}{ |c|c|c|}
			\hline
			\multicolumn{3}{|c|}{ Starobinsky inflation} \\
			\hline
			Initial conditions & \shortstack{Number of e-folds in\\the Jordan $f(R)$ frame}
				 & \shortstack{Number of e-folds\\in the Einstein frame} \\
			\hline
			$\tilde{\varphi}(0) = 9,\,\tilde{\varphi}_{\tilde{N}}(0) = -\sqrt{5.99}$   & 90.1&   86.4\\
			\hline
			$\tilde{\varphi}(0) = 5.5,\, \tilde{\varphi}_{\tilde{N}}(0) = 1$&   89.8  & 87.5\\
			\hline
			\multicolumn{3}{|c|}{Chaotic inflation} \\
			\hline
			$\tilde{\varphi}(0) = 20, \,\tilde{\varphi}_{\tilde{N}}(0) = -\sqrt{5.99}$   & 80.9&   72.7\\
			\hline
			$\tilde{\varphi}(0) = 17, \,\tilde{\varphi}_{\tilde{N}}(0) = 1$&   84.0  & 77.2\\
			\hline
		\end{tabular}
	\end{center}
	\caption{Difference in number of e-folds during inflation in both Jordan $f(R)$ frame as well in Einstein frame.}
	\label{Tab:Diff}
\end{table}

\twocolumngrid

\end{document}